\documentclass[twocolumn,floatfix]{aastex62}
\usepackage{amsmath}

\usepackage{soul}

\usepackage{graphicx}
\usepackage{times}
\usepackage[normalem]{ulem}
\usepackage{color}
\usepackage{rotating}

\begin{document}

\title{Common Envelope Wind Tunnel: The Effects of Binary Mass Ratio and Implications for the Accretion-Driven Growth of LIGO Binary Black Holes}

\correspondingauthor{Soumi De}
\email{sde101@syr.edu}

\author[0000-0002-3316-5149]{Soumi De}
\affiliation{Department of Physics, Syracuse University, Syracuse, NY 13244, USA}\affiliation{Kavli Institute for Theoretical Physics, University of California, Santa Barbara, CA 93106, USA}
\author[0000-0002-1417-8024]{Morgan MacLeod}\affiliation{Harvard-Smithsonian Center for Astrophysics, 60 Garden Street, Cambridge, MA 02138, USA}
\author[0000-0001-5256-3620]{Rosa Wallace Everson}\altaffiliation{NSF Graduate Research Fellow}\affiliation{Department  of  Astronomy  and  Astrophysics,  University  of
California, Santa Cruz, CA 95064}\affiliation{Niels Bohr Institute, University of Copenhagen, Blegdamsvej 17, 2100 Copenhagen, Denmark}
\author[0000-0003-3062-4773]{Andrea Antoni}
\affiliation{Department of Astronomy, University of California, Berkeley, CA 94720, USA}
\author[0000-0002-6134-8946]{Ilya Mandel}\affiliation{School of Physics and Astronomy, Monash University, Clayton, VIC 3800, Australia}
\affiliation{OzGrav: The ARC Centre of Excellence for Gravitational Wave Discovery, Australia} 
\affiliation{School of Physics and Astronomy, University of Birmingham, Edgbaston, Birmingham B15 2TT, United Kingdom}
\affiliation{Niels Bohr Institute, University of Copenhagen, Blegdamsvej 17, 2100 Copenhagen, Denmark}
\author[0000-0003-2558-3102]{Enrico Ramirez-Ruiz}\affiliation{Department  of  Astronomy  and  Astrophysics,  University  of
California, Santa Cruz, CA 95064}\affiliation{Niels Bohr Institute, University of Copenhagen, Blegdamsvej 17, 2100 Copenhagen, Denmark}



\begin{abstract}
We present three-dimensional local hydrodynamic simulations of flows around objects embedded within stellar envelopes using a ``wind tunnel'' formalism. Our simulations model the common envelope dynamical inspiral phase in binary star systems in terms of dimensionless flow characteristics. We present suites of simulations that study the effects of varying the binary mass ratio, stellar structure, equation of state, relative Mach number of the object's motion through the gas, and density gradients across the gravitational focusing scale. For each model, we measure coefficients of accretion and drag experienced by the embedded object. These coefficients regulate the coupled evolution of the object's masses and orbital tightening during the dynamical inspiral phase of the common envelope. We extrapolate our simulation results to accreting black holes with masses comparable to that of the population of LIGO black holes. We demonstrate that the mass and spin accrued by these black holes per unit orbital tightening are directly related to the ratio of accretion to drag coefficients. We thus infer that the mass and dimensionless spin of initially non-rotating black holes change by of order 1\% and 0.05, respectively, in a typical example scenario. Our prediction that the masses and spins of black holes remain largely unmodified by a common envelope phase aids in the interpretation of the properties of the growing observed population of merging binary black holes. Even if these black holes passed through a common envelope phase during their assembly, features of mass and spin imparted by previous evolutionary epochs should be preserved. 
\end{abstract}




\section{Introduction} \label{sec:intro}
A common envelope phase is a short episode in the life of a binary star system in which the two components of the binary evolve inside a shared envelope. Common envelope phases typically occur when one of the stars in the binary expands, engulfing its companion object \citep{Paczynski:1976,1978ApJ...222..269T,1993PASP..105.1373I,2010NewAR..54...65T,2013A&ARv..21...59I,2017PASA...34....1D}. Inside the common envelope, the embedded companion object interacts with the material flowing past it, giving rise to dynamical friction drag forces \citep{1943ApJ....97..255C,1999ApJ...513..252O}. These drag forces lead to an orbital tightening as the two objects spiral in. 
Common envelope phases are thought to be critical to the formation of compact-object binaries that subsequently merge through the emission of gravitational radiation~\citep{Heuvel:1976,Smarr:1976} \citep[see, e.g.,][for a review]{Mandel:2018hfr}. Thus, understanding the common envelope phase is important for understanding the formation channel and evolutionary history of merging compact-object binaries, such as those observed by the LIGO and Virgo gravitational-wave detectors \citep{TheLIGOScientific:2014jea, TheVirgo:2014hva}.

Significant theoretical effort has gone into modeling the physical processes of common envelope phases. This work has been challenging because of the range of physically-significant spatial and temporal scales, as well as the range of potentially important physical processes \citep{1993PASP..105.1373I,2013A&ARv..21...59I}. One crucial example is the energy release from the recombination of ionized hydrogen and helium \citep{Nandez:2015,Ivanova:2016,Lucy:1967,Roxburgh:1967,Han:1994,Han:2002}.  Efforts have often either focused on global hydrodynamic modeling of the overall encounter  \citep[for example, the recent work of][]{Ricker_2007,Passy_2011,Ricker_2012,Ohlmann:2016a,Ohlmann:2016b,Iaconi:2017,Iaconi:2018,Chamandy:2018,Chamandy:2018a,Chamandy:2019psk,Fragos:2019box}, 
or local hydrodynamic simulations that simplify and zoom in on one aspect of the larger encounter \citep[e.g.][]{Fryxell:1987,Fryxell:1988,Taam:1989,Sandquist_1998,MacLeod_2015,MacLeod:2014yda,MacLeod:2017}.
A synthesis of these approaches offers a pathway toward understanding the complex gas dynamics of common envelope phases.

This paper extends previous work on local simulations of gas flow past an object inspiraling through the gaseous surroundings of a common envelope. We use the ``wind tunnel'' formalism, first presented in \citet{MacLeod:2014yda}, and expanded in \citet{MacLeod:2017} to study the flow past a compact object embedded in the stellar envelope of a red giant or asymptotic giant branch star. The stellar profile of the donor at the onset of the dynamically unstable mass transfer depends on the mass ratio and initial separation between the centers of the two stars in the binary. 
We focus in particular on the variation in the properties as the binary mass ratio changes, and we present two suites of simulations with ideal gas equations of state characterized by adiabatic exponents $\gamma = 4/3$ and $\gamma = 5/3$, which bracket the range of typical values in stellar envelopes \citep[e.g.][]{MacLeod:2017,Murguia-Berthier:2017}.

This paper is organized as follows. In Sec.~\ref{sec:CE_params} we describe the common envelope flow parameters and conditions. We describe gravitational focusing in common envelope flows and illustrate the parameter space that controls the properties of the local flow past an object embedded in a common envelope. In Sec.~\ref{sec:hydro_sims} we describe the wind tunnel setup for hydrodynamic simulations, describe the model parameters, illustrate how the flow evolves through the simulations, and the quantities that we compute as a product of the simulations. We present hydrodynamic simulations using the wind tunnel setup for common envelope flows with a $\gamma = 4/3$ and $\gamma = 5/3$ equation of state, describe the flow characteristics, and the results obtained from the simulations. In Sec.~\ref{sec:implications}, we extrapolate our simulation results for the scenario of a black hole inspiraling through the envelope of its companion. We estimate the mass and spin accrued by black holes during the common envelope phase and derive implications for the effect of this phase on the properties of black holes in merging binaries that constitute LIGO-Virgo sources.
We conclude in Sec.~\ref{sec:conclusions}. A companion paper, \citet{Rosa:2019}, explores the validity of the expression of realistic stellar models in the dimensionless terms adopted here.

\vspace{5mm}
\section{Common envelope flow parameters and conditions}\label{sec:CE_params} 

\subsection{Characteristic Scales}
The Hoyle-Lyttleton (HL) theory of accretion ~\citep{1939PCPS...35..405H,1944MNRAS.104..273B,Edgar:2004}, is used extensively to describe accretion onto a compact object having a velocity relative to the ambient medium. We use that as a starting point to consider an embedded, accreting object of mass $M_2$ moving with velocity $v_{\infty}$ relative to a surrounding gas of unperturbed density $\rho_\infty$ that follows a stellar profile typical of a common envelope. The characteristic impact parameter inside which gas is gravitationally focused toward the embedded object and can potentially accrete is 
\begin{equation}
R_{\mathrm a} = \frac{2GM_2}{v_{\infty}^{2}},
\end{equation}
which implies a characteristic interaction cross section of $\pi R_{\mathrm{a}}^2$ \citep{1939PCPS...35..405H}. 
The corresponding mass flux through this cross section, and potential mass accretion rate in HL flows can be written as~\citep{Edgar:2004},
\begin{equation}\label{eq:mdotHL}
\dot M_{\rm HL} = \pi R_{\mathrm{a}}^2 \rho_\infty v_\infty. 
\end{equation}
The characteristic scales for momentum and energy dissipation due to gravitational interaction \citep{1999ApJ...513..252O} can be derived from this cross section as well. The characteristic scale for the momentum dissipation rate, or force, is
\begin{equation}\label{eq:fHL}
    F_{\rm HL} = \pi R_{\mathrm{a}}^2 \rho_\infty v_\infty^2 = \dot M_{\rm HL} v_\infty, 
\end{equation}
and the characteristic energy dissipation rate is
\begin{equation}
    \dot E_{\rm HL} = \pi R_{\mathrm{a}}^2 \rho_\infty v_\infty^3 = \dot M_{\rm HL} v_\infty^2, 
\end{equation}
if we assume that all momentum and energy passing through the interaction cross section $\pi R_{\mathrm{a}}^2$ are dissipated. 

\subsection{Common Envelope Parameters}

We imagine that the embedded object $M_2$ is spiralling in to tighter orbital separations within the envelope of a giant-star primary.  The core of the primary is fixed at $r = 0$ and the orbital radius of $M_2$ within  the primary's envelope is $r = a$. Thus, the stellar cores are separated by a distance $a$, smaller than the original radius of the primary.  We use $M_1(r)$ to denote the mass of the primary that is enclosed by the orbit of $M_2$. Therefore, the Keplerian orbital velocity is $v_{\rm k} = \sqrt{GM/a}$,
where $M = M_{1}(a) + M_{2}$ is the total enclosed mass of the binary (mass outside of the orbital separation $a$ does not contribute to the orbital velocity). The relative velocity of the secondary to the envelope gas, $v_{\infty}$, is related to the Keplerian velocity of the secondary as $v_{\infty} = f_{\rm k}v_{k}$. Thus, $f_{\rm k}$ is the fraction of the Keplerian velocity that contributes to the relative velocity. In our simulations, we adopt the simplification $f_{\rm k} = 1$. However, $f_{\rm k} < 1.0$ is possible if the orbital motion of the embedded object is partially synchronized to the donor's envelope.

\begin{figure}[t]
  \includegraphics[width=\columnwidth]{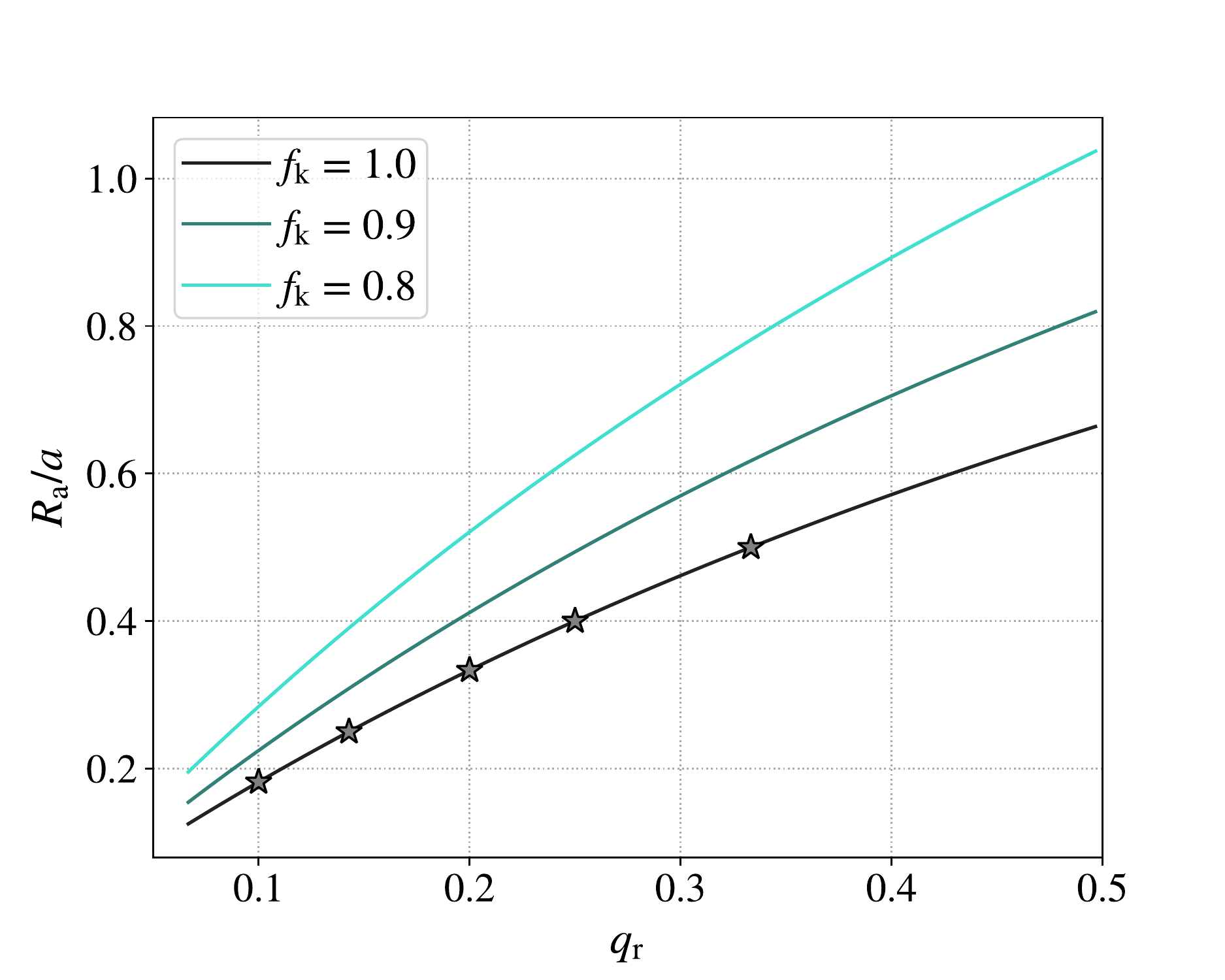}
 \caption{Fraction of the orbital separation falling within the gravitational focusing radius of the embedded object, $R_{\rm a}/a$, as a function of the binary mass ratio $q_{\rm r}$. The plot shows the relation for $f_{\rm k} = 1.0, 0.9, 0.8$, where $f_{\rm k}$ is the fraction of the Keplerian velocity contributing to the relative velocity. Markers show the points for which hydrodynamical simulations have been performed in this paper. For small mass ratios, the accretion radius of $M_2$ is small relative to the orbital separation.  When $q_{\rm r}$ is large, $R_a$ sweeps out a significant fraction of the orbital separation.  For fixed $M_2$, $q_{\rm r}$  increases as the  embedded object spirals further into the envelope of the primary.
 \label{fig:ra-q}}
\vspace{5mm}
\end{figure}

Given a relative velocity set by the orbital motion, the ratio of the gravitational focusing scale, $R_{\rm a}$, to the orbital separation, $a$, is \citep{MacLeod:2017}
\begin{equation}
\frac{R_{\mathrm a}}{a} = \frac{2}{f_{\rm k}^{2}}\frac{M_{2}}{M} = \frac{2}{f_{\rm k}^{2}}\frac{1}{1 + q_{\rm r}^{-1}},\label{eq:Ra_asfunc_a}
\end{equation}
where $q_{\rm r}=M_{2}/M_{1}(r)$ is the mass ratio between the embedded object and the mass enclosed by its orbit. Therefore, for a given value of $q_{\rm r}$, one can calculate $R_{\mathrm a}$ in terms of $a$. The variation of $R_{\mathrm a}$ in terms of $a$ with $q_{\rm r}$ is shown in Figure~\ref{fig:ra-q} for $f_{\rm k}=1, 0.9, 0.8$. As $q_{\rm r}$ increases, $R_{\mathrm a}/a$ also increases, and gives an approximate scale for the fraction of the envelope interior upon which the embedded object actively impinges.

The HL formalism assumes a homogeneous background for the embedded object. In practice, such a situation does not arise in common envelope encounters. As demonstrated in Figure \ref{fig:ra-q},  $R_{\mathrm{a}}/a$ can be a large fraction of unity for typical mass ratios. Therefore, the gaseous medium with which the embedded object interacts spans a range of densities and temperatures \citep{MacLeod_2015}. 

The flow upstream Mach number is the ratio of orbital velocity to sound speed, 
\begin{equation}
\mathcal{M}_\infty = \frac{v_\infty}{c_{s, \infty}},
\end{equation}
where we specify $c_{s, \infty}$ to be the sound speed measured at radius $r=a$ within the common envelope gas. 
Furthermore, the density gradient in stellar profiles can be expressed in terms of a local density scale height at the location of the embedded object as
\begin{equation}
H_\rho = -\rho \frac{dr}{d\rho} . 
\end{equation}
The number of scale heights encompassed by the accretion radius is then quantified by the ratio
\begin{equation}
\epsilon_\rho = \frac{R_{\mathrm{a}} }{ H_\rho} ,    
\end{equation}
which is, like other quantities, evaluated at the location of the embedded object. This density gradient breaks the symmetry of the flow envisioned in the HL scenario and gives the flow a net angular momentum relative to the accreting object \citep{MacLeod_2015,MacLeod:2014yda,MacLeod:2017,Murguia-Berthier:2017}.

\citet{MacLeod:2017} showed that there is a clear relation between Mach number and density gradient for typical common envelope flows when the (local) envelope structure is approximated as a polytrope with index 
\begin{equation}
\Gamma_{\rm s} = \bigg(\frac{d\ln{P}}{d\ln{\rho}}\bigg)_{\rm envelope}.
\end{equation}
Under the simplification of an ideal gas equation of state with adiabatic index $\gamma$, we can rewrite the hydrostatic equilibrium condition of the envelope as a relationship between $\mathcal{M}_\infty$ and $\epsilon_\rho$ \citep[Equation 18 of][]{MacLeod:2017}, 
\begin{equation}\label{eq:mach-erho}
\mathcal{M}^2_\infty = \epsilon_\rho \frac{(1 + q_{\rm r})^2}{2q_{\rm r}}f_{\rm k}^4 \bigg(\frac{\Gamma_s}{\gamma}\bigg).
\end{equation}
Thus, not all parameter combinations of $\mathcal{M}_\infty$ and $\epsilon_\rho$ are realized in common envelope phases. Instead, typical parameter combinations are $f_{\rm k}$ and $q_{\rm r}$. The validity of this approximation in the context of detailed stellar evolution models  is  discussed  in \citet{Rosa:2019}, in which it is shown that the simulations presented in this paper are still applicable to detailed stellar models described by a realistic equation of state.

\vspace{0.5cm}
\section{Hydrodynamic Simulations\label{sec:hydro_sims}}
In this section we describe hydrodynamic simulations in the Common Envelope Wind Tunnel formalism \citep{MacLeod:2017} that explore the effects of varying the binary mass ratio on coefficients of drag and accretion realized during the dynamical inspiral of an object through the envelope of its companion.  

\subsection{Numerical Method}\label{sec:method}
The Common Envelope Wind Tunnel model used in this work is a hydrodynamic setup using the FLASH Adaptive Mesh Refinement hydrodynamics code \citep{Fryxell2000}. A full description of the model is given in Section 3 of \citet{MacLeod:2017}. The basic premise is that the complex geometry of a full common envelope scenario is replaced with a 3D Cartesian wind tunnel surrounding a hypothetical embedded object. Flow moves past the embedded object and we are able to measure rates of mass accretion and drag forces. 

In the Common Envelope Wind Tunnel, flows are injected from the $-x$ boundary of the computational domain past a gravitating point mass, located at the coordinate origin of the three-dimensional domain.  To simulate accretion, the point mass is surrounded by a low pressure ``sink" of radius $R_{\rm s}$. The gas obeys an ideal gas equation of state $P=(\gamma-1)\rho e$, where $e$ is the specific internal energy. The profile of inflowing material is defined by its upstream Mach number, $\mathcal M_\infty$, and the ratio of the accretion radius to the density scale height, $\epsilon_\rho$.  Calculations are performed in code units $R_{\mathrm{a}} = v_\infty = \rho_\infty = 1$. Here $\rho_\infty$ is the density of the unperturbed profile at the location of the embedded object. This gives a time unit of $R_{\mathrm{a}}/v_\infty = 1$, which is the time taken by the flow to cross the accretion radius. The binary separation $a$ in code units is
\begin{equation}\label{eq:sepcode}
\frac{a}{R_{\rm a}} = \frac{1}{2}f_{\rm k}^2 (1 + q_{\rm r}^{-1}) .
\end{equation}
The density profile of the gas in the $\hat y$-direction is that of a polytrope with index $\Gamma_s$ in hydrostatic equilibrium with a gravitational force 
\begin{equation}
\vec{a}_{\rm grav, 1} = -\frac{GM_1(r)}{(y - y_1)^2}\hat{y},
\end{equation}
that represents the gravitational force from the primary star's enclosed mass, $M_1 (r)$. The density scale height, sound speed and upstream Mach number vary across this profile as they would in a polytropic star.  At the $+y$ and $\pm z$ boundaries, a ``diode'' boundary condition is applied, that allows material to leave but not enter into the domain.

The size of the domain is set by the mass ratio of the binary system and the effective size of the binary orbit, as described by Equation \eqref{eq:sepcode}. Gravitationally focused gas flows are sensitive to the distance over which they converge, and the size of the wake that they leave \citep[e.g.][]{1999ApJ...513..252O}. In varying the binary mass ratio, it is important to capture this physical property of differing ratio of the gravitational focus radius to the physical size of the system, equation \eqref{eq:sepcode}. In order to capture the full flow, our domain has a diameter equal to the binary separation $a$, implying that it extends a distance $\pm a/2 =  (1 + q_{\rm r}^{-1}) R_{\mathrm{a}}/4$ about the origin in the $\pm x$, $\pm y$, and $\pm z$ directions. 
  
This domain is spatially resolved by  cubic blocks that have extent of $R_{\rm a}/2$ in each direction, and each block is made of $8^3$ zones. The largest zones have length $R_{\mathrm{a}}/16$. We allow for five levels of adaptive mesh refinement, giving the smallest zones length $R_{\mathrm{a}}/256$. We enforce maximum refinement around the embedded object at all times.

\subsection{Model Parameters}

The simulations that we present later in this section assume $\Gamma_s = \gamma$ and $f_{\rm k} = 1$. We are therefore modeling constant entropy stellar envelope material (as in a convective envelope of a giant star) and relative velocities between the embedded object and the background gas equal to the Keplerian velocity. All models adopt a sink radius for measuring accretion of $R_{\rm s} = 0.05R_{\mathrm{a}}$ around the embedded object. In Section \ref{sec:sink}, we perform simulations with varying sink radius and we discuss the dependence of our results on this parameter. 

\begin{figure}[t]
  \includegraphics[width=\columnwidth]{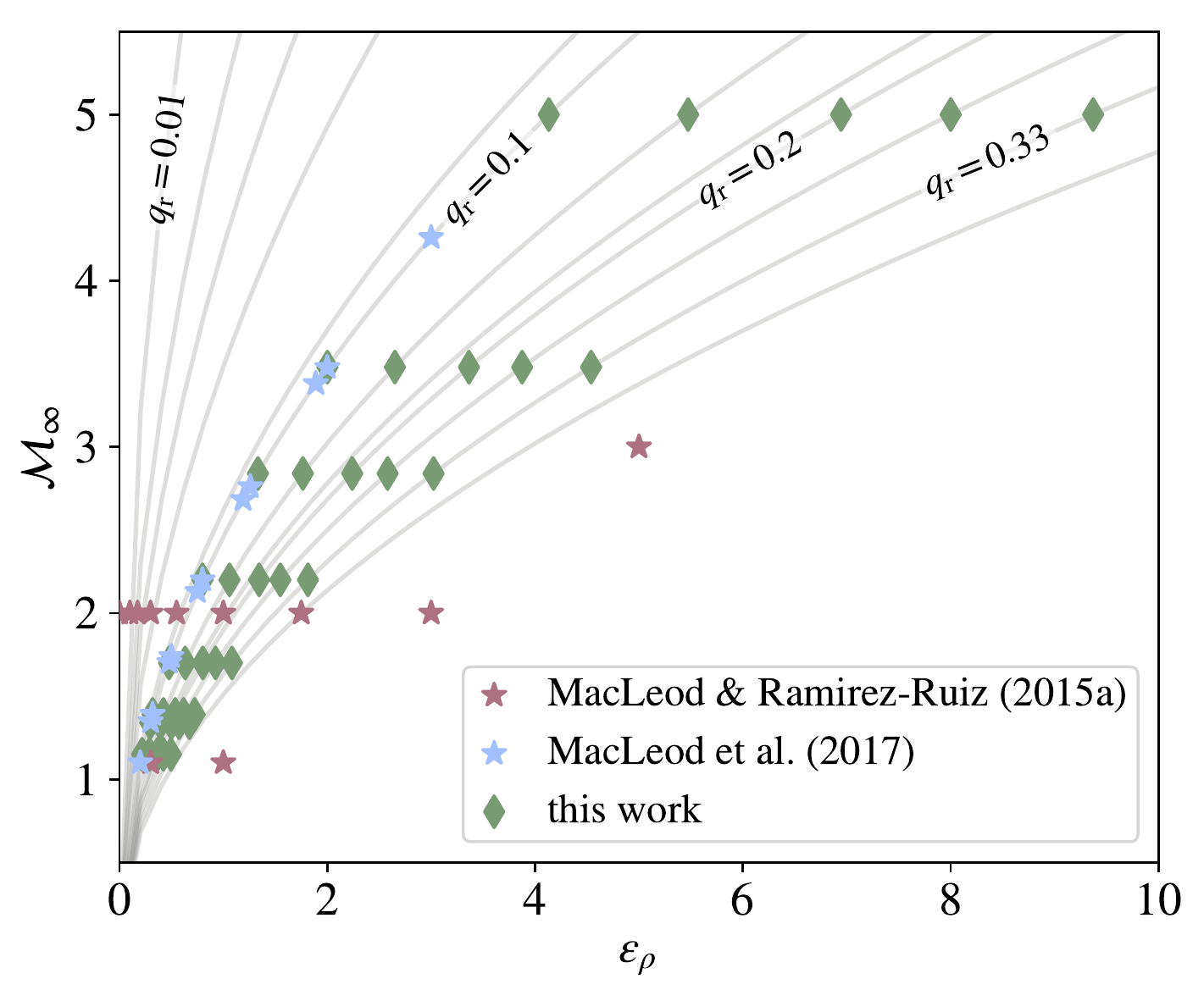}
 \caption{Points in the $\mathcal{M}_{\infty}$-$\epsilon_{\rho}$ space representing flow parameters for simulations performed with the ``wind tunnel'' setup. For polytropic envelopes, each combination of $\mathcal{M}_{\infty}$ and $\epsilon_{\rho}$ has a corresponding $q_{\rm r}$ value \citep{MacLeod:2017}. Simulations are shown on lines of constant $q_{\rm r}$, with the exception of three simulations from \citet{MacLeod_2015} that do not follow the polytropic relation. The simulations in this work expand upon the previous work as labeled, extending across both axes to higher Mach numbers and steeper density gradients, significantly extending coverage across the region of parameter space realized in realistic stellar profiles as detailed in \citet{Rosa:2019}.\label{fig:parameterspace}}
\vspace*{5mm}
\end{figure}

This leaves three flow parameters in equation~\eqref{eq:mach-erho}: $\mathcal{M}_\infty$, $\epsilon_\rho$, and $q_{\rm r}$, only two of which can be chosen independently. Figure \ref{fig:parameterspace} shows the simulation grid presented in this paper and those in \citet{MacLeod:2014yda,MacLeod:2017} in the $\mathcal{M}_\infty - \epsilon_\rho$ space. The simulations in this paper expand the parameter space covered in the previous papers with a broader range of $\mathcal{M}_\infty$ (therefore $\epsilon_\rho$) and, crucially, models of varying mass ratio, $q_{\rm r}$. We construct a grid of $q_{\rm r} - \mathcal{M}_\infty$ values, with $q_{\rm r}$ values $1/10$, $1/7$, $1/5$, $1/4$, $1/3$. For each value of $q_{\rm r}$ we perform simulations with $\mathcal{M}_\infty$ of $1.15$, $1.39$, $1.69$, $2.2$, $2.84$, $3.48$, and $5.0$. It was shown in ~\citet{MacLeod_2015} with the help of MESA simulations of 1-16$M_\odot$ stars evolved from the zero-age main sequence to the giant branch expansion, that typical upstream Mach number values range from $\mathcal{M}_\infty~\approx 2$ in the deep interior to $\mathcal{M}_\infty \gtrsim 5$ near the stellar limb. Extending these results in \citet{Rosa:2019}, MESA is used to evolve a broader range of stellar masses 3--90$M_\odot$ with binary mass ratios of 0.1--0.35, finding $1.5 \lesssim \mathcal{M}_\infty < 7$ in giant branch stellar envelopes.

Tabulated model parameters are presented in Table~\ref{tab:sims_43_params} and Table~\ref{tab:sims_53_params}. We divide our discussion in the subsequent sections to consider the $\gamma = \Gamma_{\rm s}=4/3$ and $\gamma=\Gamma_{\rm s} = 5/3$ models separately.

\begin{table}[t]
\centering\begin{tabular}{lcccccc} 
\hline
\rule{0pt}{3ex}
Name & $\gamma$ & $q_{\rm r}$ & $\mathcal{M}_\infty$ & $\epsilon_\rho$ & $C_{\mathrm a}$ & $C_{\rm d}$ \\
\hline\hline
\rule{0pt}{3ex}%
\vspace*{0.1cm}
A1 & 4/3 & 0.1 & 1.15 & 0.22 & 0.70 & 1.20 \\
A2 & 4/3 & 0.1 & 1.39 & 0.32 & 0.77 & 1.44 \\
A3 & 4/3 & 0.1 & 1.69 & 0.47 & 0.66 & 1.60 \\
A4 & 4/3 & 0.1 & 2.20 & 0.80 & 0.38 & 1.91 \\
A5 & 4/3 & 0.1 & 2.84 & 1.33 & 0.10 & 3.36 \\
A6 & 4/3 & 0.1 & 3.48 & 2.00 & 0.07 & 5.44 \\
A7 & 4/3 & 0.1 & 5.00 & 4.13 & 0.04 & 18.92 \\
A8 & 4/3 & 0.143 & 1.15 & 0.29 & 0.74 & 1.03 \\
A9 & 4/3 & 0.143 & 1.39 & 0.42 & 0.65 & 1.20 \\
A10 & 4/3 & 0.143 & 1.70 & 0.63 & 0.52 & 1.22 \\
A11 & 4/3 & 0.143 & 2.20 & 1.06 & 0.26 & 1.41 \\
A12 & 4/3 & 0.143 & 2.84 & 1.77 & 0.09 & 2.93 \\
A13 & 4/3 & 0.143 & 3.48 & 2.65 & 0.10 & 5.15 \\
A14 & 4/3 & 0.143 & 5.00 & 5.47 & 0.07 & 19.38 \\
A15 & 4/3 & 0.2 & 1.15 & 0.37 & 0.80 & 0.80 \\
A16 & 4/3 & 0.2 & 1.39 & 0.54 & 0.76 & 1.01 \\
A17 & 4/3 & 0.2 & 1.70 & 0.80 & 0.45 & 0.97 \\
A18 & 4/3 & 0.2 & 2.20 & 1.34 & 0.22 & 1.05 \\
A19 & 4/3 & 0.2 & 2.84 & 2.24 & 0.11 & 2.02 \\
A20 & 4/3 & 0.2 & 3.48 & 3.36 & 0.09 & 4.34 \\
A21 & 4/3 & 0.2 & 5.00 & 6.94 & 0.29 & 12.93 \\
A22 & 4/3 & 0.25 & 1.15 & 0.42 & 0.79 & 0.65 \\
A23 & 4/3 & 0.25 & 1.39 & 0.62 & 0.74 & 0.83 \\
A24 & 4/3 & 0.25 & 1.70 & 0.93 & 0.38 & 0.82 \\
A25 & 4/3 & 0.25 & 2.20 & 1.55 & 0.23 & 0.85 \\
A26 & 4/3 & 0.25 & 2.84 & 2.58 & 0.13 & 1.66 \\
A27 & 4/3 & 0.25 & 3.48 & 3.87 & 0.13 & 3.11 \\
A28 & 4/3 & 0.25 & 5.00 & 8.00 & 0.61 & 7.73 \\
A29 & 4/3 & 0.3333 & 1.15 & 0.50 & 0.64 & 0.53 \\
A30 & 4/3 & 0.3333 & 1.39 & 0.73 & 0.62 & 0.65 \\
A31 & 4/3 & 0.3333 & 1.70 & 1.08 & 0.37 & 0.61 \\
A32 & 4/3 & 0.3333 & 2.20 & 1.81 & 0.23 & 0.65 \\
A33 & 4/3 & 0.3333 & 2.84 & 3.02 & 0.13 & 1.25 \\
A34 & 4/3 & 0.3333 & 3.48 & 4.54 & 0.18 & 1.91 \\
A35 & 4/3 & 0.3333 & 5.00 & 9.37 & 1.06 & 5.28 \\
\hline
\end{tabular}
\caption{Input parameters---$q_{\rm r}$, $\mathcal {M}_\infty$, $\epsilon_\rho$ and results---$C_{\mathrm a}$, $C_{\mathrm d}$ for $\gamma = 4/3$ simulations. The $C_\mathrm{a}$, $C_\mathrm{d}$ entries are median values computed over simulation times $10~R_{\rm a} / v_\infty < t < 30~R{\rm a} / v_\infty$.}
\label{tab:sims_43_params}
\end{table}

\begin{table}[t]
\centering\begin{tabular}{lcccccc} 
\hline
\rule{0pt}{3ex}
Name & $\gamma$ & $q_{\rm r}$ & $\mathcal{M}_\infty$ & $\epsilon_\rho$ & $C_{\mathrm a}$ & $C_{\rm d}$ \\
\hline\hline
\rule{0pt}{3ex}%
\vspace*{0.1cm}
B1 & 5/3 & 0.1 & 1.15 & 0.22 & 0.36 & 0.79 \\
B2 & 5/3 & 0.1 & 1.39 & 0.32 & 0.38 & 0.95 \\
B3 & 5/3 & 0.1 & 1.69 & 0.47 & 0.21 & 0.99 \\
B4 & 5/3 & 0.1 & 2.20 & 0.80 & 0.14 & 1.35 \\
B5 & 5/3 & 0.1 & 2.84 & 1.33 & 0.05 & 2.07 \\
B6 & 5/3 & 0.1 & 3.48 & 2.00 & 0.02 & 3.03 \\
B7 & 5/3 & 0.1 & 5.00 & 4.13 & 0.01 & 6.22 \\
B8 & 5/3 & 0.143 & 1.15 & 0.29 & 0.36 & 0.58 \\
B9 & 5/3 & 0.143 & 1.39 & 0.42 & 0.35 & 0.79 \\
B10 & 5/3 & 0.143 & 1.70 & 0.63 & 0.24 & 0.85 \\
B11 & 5/3 & 0.143 & 2.20 & 1.06 & 0.13 & 1.14 \\
B12 & 5/3 & 0.143 & 2.84 & 1.77 & 0.05 & 1.63 \\
B13 & 5/3 & 0.143 & 3.48 & 2.65 & 0.03 & 2.42 \\
B14 & 5/3 & 0.143 & 5.00 & 5.47 & 0.03 & 5.70 \\
B15 & 5/3 & 0.2 & 1.15 & 0.37 & 0.38 & 0.40 \\
B16 & 5/3 & 0.2 & 1.39 & 0.54 & 0.37 & 0.57 \\
B17 & 5/3 & 0.2 & 1.70 & 0.80 & 0.22 & 0.65 \\
B18 & 5/3 & 0.2 & 2.20 & 1.34 & 0.13 & 0.84 \\
B19 & 5/3 & 0.2 & 2.84 & 2.24 & 0.06 & 1.24 \\
B20 & 5/3 & 0.2 & 3.48 & 3.36 & 0.06 & 1.85 \\
B21 & 5/3 & 0.2 & 5.00 & 6.94 & 0.04 & 4.76 \\
B22 & 5/3 & 0.25 & 1.15 & 0.42 & 0.39 & 0.32 \\
B23 & 5/3 & 0.25 & 1.39 & 0.62 & 0.39 & 0.46 \\
B24 & 5/3 & 0.25 & 1.70 & 0.93 & 0.20 & 0.54 \\
B25 & 5/3 & 0.25 & 2.20 & 1.55 & 0.09 & 0.65 \\
B26 & 5/3 & 0.25 & 2.84 & 2.58 & 0.07 & 1.03 \\
B27 & 5/3 & 0.25 & 3.48 & 3.87 & 0.07 & 1.54 \\
B28 & 5/3 & 0.25 & 5.00 & 8.00 & 0.11 & 3.55 \\
B29 & 5/3 & 0.3333 & 1.15 & 0.50 & 0.42 & 0.17 \\
B30 & 5/3 & 0.3333 & 1.39 & 0.73 & 0.35 & 0.31 \\
B31 & 5/3 & 0.3333 & 1.70 & 1.08 & 0.21 & 0.42 \\
B32 & 5/3 & 0.3333 & 2.20 & 1.81 & 0.10 & 0.50 \\
B33 & 5/3 & 0.3333 & 2.84 & 3.02 & 0.08 & 0.80 \\
B34 & 5/3 & 0.3333 & 3.48 & 4.54 & 0.09 & 1.24 \\
B35 & 5/3 & 0.3333 & 5.00 & 9.37 & 0.15 & 2.76 \\
\hline
\end{tabular}
\caption{Input parameters---$q_{\rm r}$, $\mathcal M_\infty$, $\epsilon_\rho$ and results---$C_{\mathrm a}$, $C_{\mathrm d}$ for $\gamma = 5/3$ simulations. The $C_\mathrm{a}$, $C_\mathrm{d}$ entries are median values computed over simulation times $10 \, R_a / v_\infty < t < 30$ $R_a / v_\infty$.}
\label{tab:sims_53_params}
\end{table}

\subsection{Model Time Evolution and Diagnostics}

\begin{figure}[t]
  \includegraphics[width=\columnwidth]{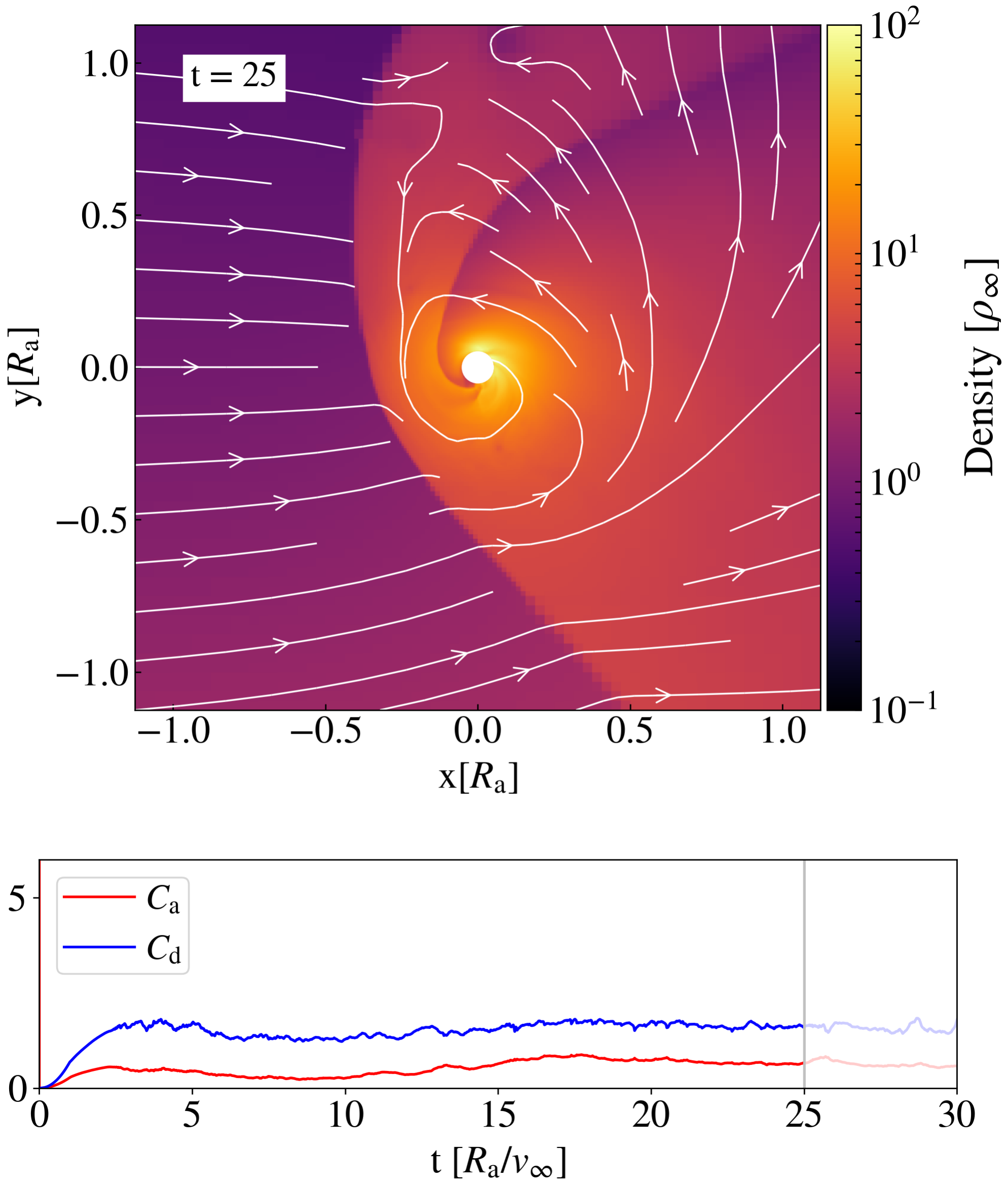}
 \caption{Movie of simulation performed with ideal gas equation of state adiabatic constant $\gamma = 4/3$, mass ratio $q_{\rm r} = 0.1$, and upstream Mach number $\mathcal{M}_\infty = 1.69$ in the ``wind tunnel'' setup. The top panel shows the formation of the shock and the evolution of the flow past the compact object embedded in the envelope of its companion star. Plotted is the density in units of $\rho_\infty$ in the orbital ($x$-$y$) plane of the binary, with the white dot at the coordinate origin representing the embedded companion object. The lines with arrowheads in white represent streamlines following the velocity field in the flow. The bottom panel shows the time series of coefficients of accretion $C_{\mathrm a}$ (in red) and drag $C_{\mathrm d}$ (blue) for the full simulation. The gray vertical line tracks the instantaneous $C_{\mathrm a}$, $C_{\mathrm d}$ values as the simulation progresses. The time quoted in the movie is in code units $R_{\mathrm{a}}/v_\infty$, where $R_{\rm a}$ is the accretion radius and $v_\infty$ is the relative velocity of the flow past the embedded object. \label{fig:video_sim}}  
\vspace*{5mm}
\end{figure}

In Figure \ref{fig:video_sim} (animated version online) we show the time evolution of a representative model (A3) with parameters $\gamma = 4/3$, $q_{\rm r} = 1/10$, and  $\mathcal{M}_\infty = 1.69$. The top panel in Figure \ref{fig:video_sim} shows a slice through the orbital ($z = 0$) plane of the binary with the white dot at the origin representing the absorbing sink around the embedded companion object. We show a section of the computational domain extending between $\pm R_{\rm a}$. The full domain extends between $\pm (1 + q_{\rm r}^{-1}) R_{\mathrm{a}}/4 = \pm 2.75 R_{\rm a}$ in each direction. The background gas injected into the domain at the $-x$ boundary, with speed $\mathcal{M}_\infty$, carries with it the density profile set by $\epsilon_\rho$ (the center of the primary is located at $y = - a$, so the density increases with decreasing $y$). Once material enters the domain, it is gravitationally focused by the embedded object and a bow shock forms due to the supersonic motion of the embedded object relative to the gas. Denser material is drawn in from deeper within the star $y<0$, such that asymmetry is introduced into the bow shock, and net rotation is imparted into the post-shock flow \citep{MacLeod_2015,MacLeod:2017}. While most of the injected material exits the domain through the $+x$ and $+y$ boundaries, some is accreted into the central sink.

As the simulation progresses, we monitor rates of mass and momentum accretion into the central sink \citep[equations 24 and 25 of][]{MacLeod:2017}, as well as the gaseous dynamical friction drag force that arises from the overdensity in the wake of the embedded object \citep[equation 28 of][]{MacLeod:2017}. 
We define the coefficients of accretion and drag to be the multiple of their corresponding HL 
values, equations~\eqref{eq:mdotHL} and \eqref{eq:fHL}, respectively, realized in our simulations.  That is, the coefficient of accretion is
\begin{equation}\label{eq:C_a}
    C_{\mathrm a} = \frac{\dot M}{\pi R_{\mathrm{a}}^2 \rho_\infty v_\infty} = \frac{\dot M }{ M_{\rm{HL}}}
\end{equation}
where $\dot M$ is the mass accretion rate measured in the simulation. The coefficient of drag is 
\begin{equation}\label{eq:C_d}
    C_{\mathrm d} = \frac{F_{\mathrm{df}} + F_{\dot p_x} }{ \pi R_{\mathrm{a}}^2 \rho_\infty v_\infty^2} = \frac{F_{\rm d} }{ F_{\rm{HL}}},
\end{equation}

where $F_{\mathrm{df}}$ is the dynamical friction drag force, $F_{\dot p_x}$ is the force due to linear momentum accretion, and $F_{\rm d} = F_{\mathrm{df}} + F_{\dot p_x}$ is the net drag force acting on the embedded object due to the gas. $F_{\mathrm{df}}$ is computed by performing a volume integral over the spherical shell of inner radius $R_{\rm s}$ and outer radius $(1 + q_{\rm r}^{-1}) R_{\mathrm{a}}/4$ (the size of the computational domain in the $\pm~x, \pm~y, \pm~z$ directions).
The bottom panel of Figure \ref{fig:video_sim} shows $C_{\mathrm a}$ and $C_{\mathrm d}$ as a function of time for model A3. We run our simulations for a duration $t = 30~R_{\rm a}/v_\infty$ (ie. 30~$\times$ code units). The flow sets up during an initial transient phase, which is $\approx 8~R_{\rm a}/v_\infty$ for model A3 presented in Figure~\ref{fig:video_sim}, after which the rates of accretion and drag subside to relatively stable values. The upstream density gradient imparts turbulence to the flow, which introduces a chaotic time variability to the accretion rate and drag. Therefore, we report median values of the $C_{\rm a}$ and $C_{\rm d}$ time series from the steady state duration of the flow, $10 \, R_a / v_\infty < t < 30$ $R_a / v_\infty$ in the remainder of the paper, though $C_{\rm a}$ and $C_{\rm d}$ are typically close to their steady-state values after a time $a/v_\infty$. 

Recently, \citet{Chamandy:2019psk} has undertaken a detailed analysis of forces in their global models of common envelope phases. One of their findings is that during the dynamical inspiral phase, flow properties and forces are very similar to those realized in local simulations such as those presented here. For example, Figure \ref{fig:video_sim} is very similar to \citet{Chamandy:2019psk}'s Figure 7.

\vspace*{2.5mm}
\subsection{Gas Flow}\label{sec:hydro}
In this section, we discuss the properties and morphology of gas flow in our Common Envelope Wind Tunnel experiments for the models tabulated in Tables \ref{tab:sims_43_params} and \ref{tab:sims_53_params}. We focus, in particular, on the differences that arise as we vary the dimensionless characteristics of the flow in the form of upstream Mach number, mass ratio, and gas adiabatic index. 

\subsubsection{Dependence on Mach Number, ${\cal M}_\infty$}\label{sec:hydro_mach}

\begin{figure*}[t]
\centering
\includegraphics[width=16.5cm]{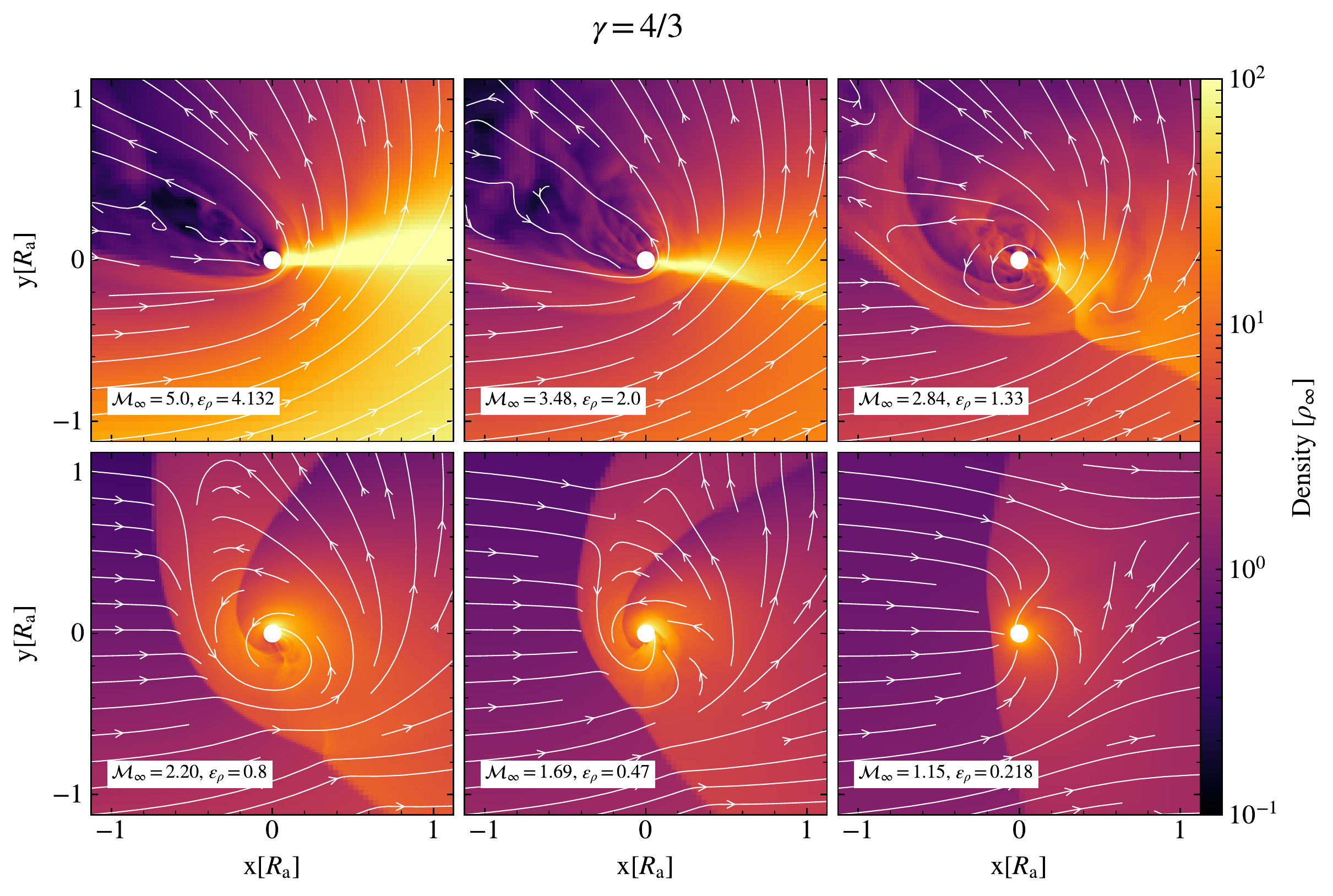}
\includegraphics[width=16.5cm]{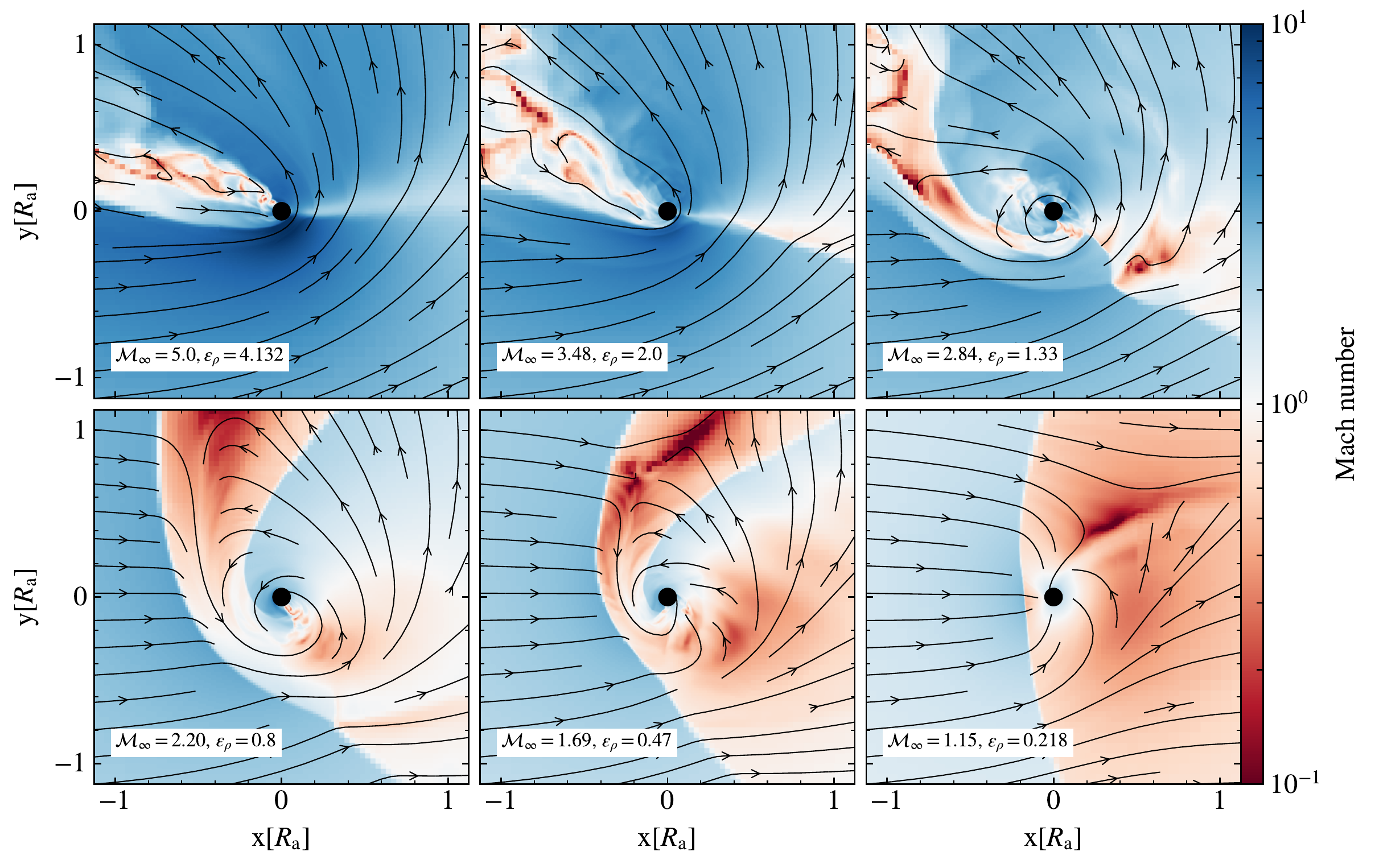}
\caption{Slices of density in units of $\rho_\infty$ (upper panels) and Mach number (lower panels) through the orbital ($x$-$y$) plane, for a fixed mass ratio $q_{\rm r}$ and varying upstream Mach number $\mathcal{M}_\infty$, for the simulation suite $(\Gamma_{\rm s}, \gamma) = (4/3, 4/3)$. The simulations use $q_{\rm r} = 0.1$ and $\mathcal{M}_\infty$ 5.0, 3.48, 2.84, 2.20, 1.69, and 1.15 corresponding to density gradients $\epsilon_\rho$ of 4.132, 2.0, 1.33, 0.8, 0.47, and 0.218 respectively. The slices compare the state of the flow at simulation time $t = 30~R_{\mathrm{a}}/v_\infty$. Moving from the highest to the lowest $\mathcal{M}_\infty$, the slices show the pattern of the flow around the embedded companion object as it inspirals from the outer to the inner regions of the primary star's envelope.  
\label{fig:sims_g43_fix_q_vary_mach}}
\end{figure*}

\begin{figure*}[t]
\centering
\includegraphics[width=16.5cm]{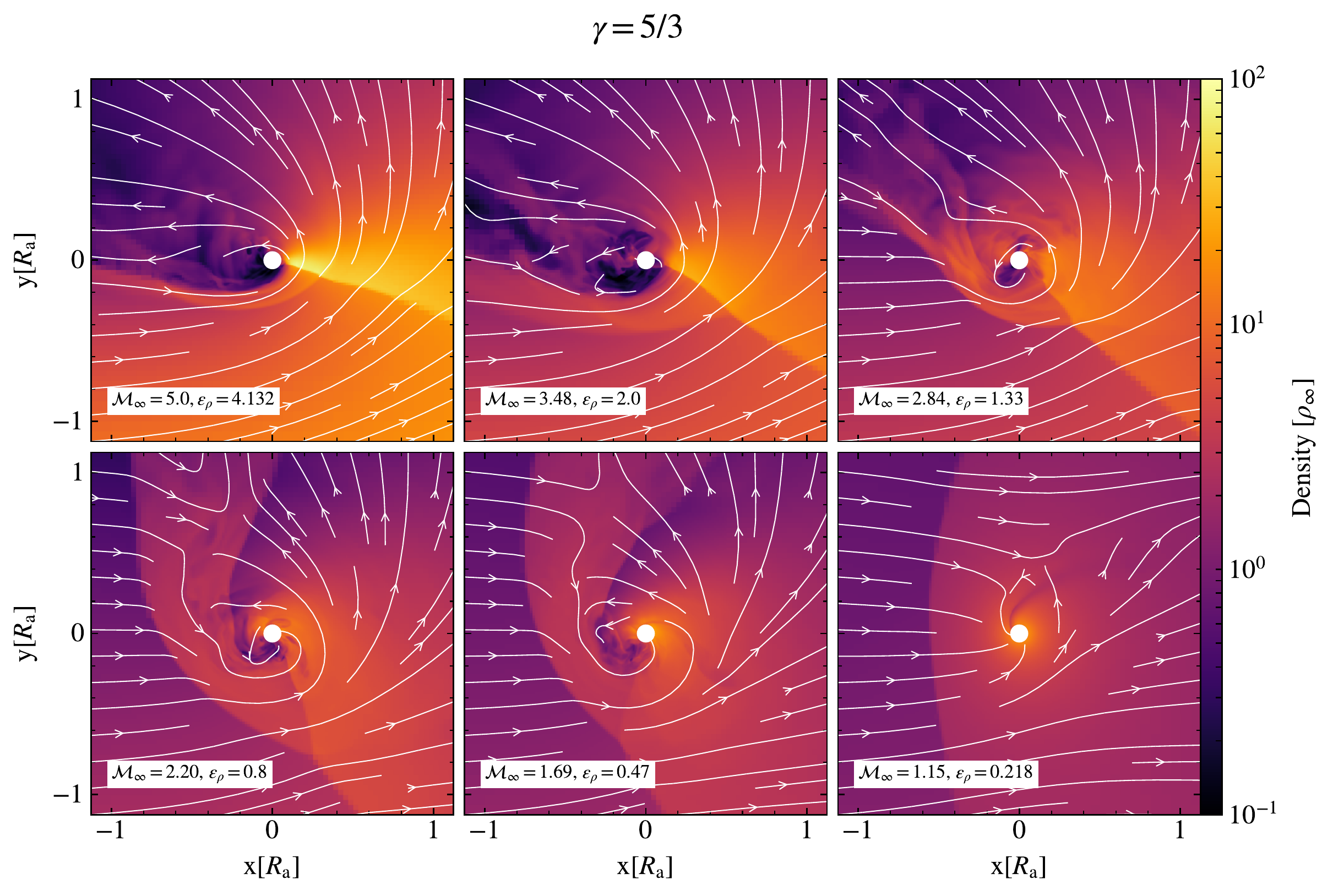}
\includegraphics[width=16.5cm]{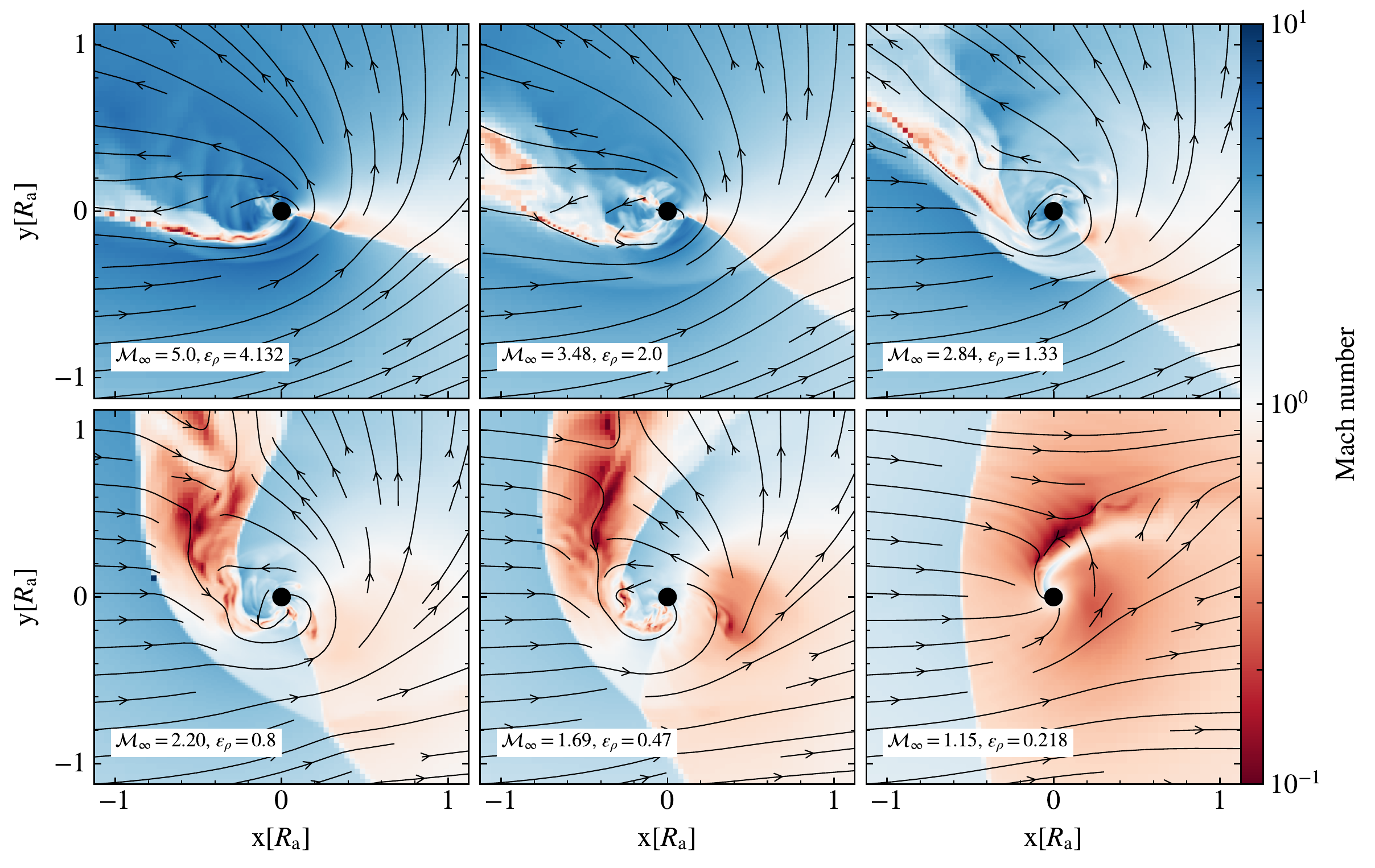}
\caption{Slices of density in units of $\rho_\infty$ (upper panels) and Mach number (lower panels) through the orbital ($x$-$y$) plane, for a fixed mass ratio $q_{\rm r}$ and varying upstream Mach number $\mathcal{M}_\infty$, for the simulation suite $(\Gamma_{\mathrm s}, \gamma) = (5/3, 5/3)$. The simulations use $q_{\rm r} = 0.1$ and $\mathcal{M}_\infty$ 5.0, 3.48, 2.84, 2.20, 1.69, and 1.15 corresponding to density gradients $\epsilon_\rho$ of 4.132, 2.0, 1.33, 0.8, 0.47, and 0.218 respectively. The slices compare the state of the flow at simulation time $t = 30~R_{\mathrm{a}}/v_\infty$. Moving from the highest to the lowest $\mathcal{M}_\infty$, the slices show the pattern of the flow around the embedded companion object as it inspirals from the outer to the inner regions of the primary star's envelope. \label{fig:sims_g53_fix_q_vary_mach}}
\end{figure*}

Figures \ref{fig:sims_g43_fix_q_vary_mach} and \ref{fig:sims_g53_fix_q_vary_mach} show slices of density and Mach number through the orbital ($x$-$y$) plane from the models with $q_{\rm r} = 1/10$ and a range of $\mathcal{M}_\infty$ and corresponding $\epsilon_\rho$ values. In Figure \ref{fig:sims_g43_fix_q_vary_mach} models from Table \ref{tab:sims_43_params} are presented, which have $\gamma = \Gamma_{\mathrm s} = 4/3$, while in Figure \ref{fig:sims_g53_fix_q_vary_mach} models from Table \ref{tab:sims_53_params} are presented, which have $\gamma = \Gamma_{\mathrm s} = 5/3$. In these slices, the $x$ and $y$ axes show distances in units of the accretion radius $R_{\mathrm{a}}$, and we overplot streamlines of the velocity field within the $x$-$y$ plane.

Higher Mach numbers imply steeper density gradients relative to the accretion radius, following equation \eqref{eq:mach-erho}.  These conditions tend to be found in the outer regions of the stellar envelope, whereas lower Mach numbers and shallower density gradients are more representative of flows found deeper in the stellar envelope. Thus, the sequence of Mach numbers approximates the inspiral of an object from the outer regions of the envelope of the donor star toward its center.

Figures \ref{fig:sims_g43_fix_q_vary_mach} and \ref{fig:sims_g53_fix_q_vary_mach} demonstrate how a decreasing $\mathcal{M}_\infty$ for fixed $q_{\rm r}$ affects the flow characteristics. A key distinction is that the flow symmetry is more dramatically broken at high $\mathcal{M}_\infty$ (and $\epsilon_{\rho}$), and it gradually becomes more symmetric with decreasing $\mathcal{M}_\infty$ and $\epsilon_{\rho}$ \citep{MacLeod_2015,MacLeod:2017}. It is important to emphasize that the controlling parameter generating this asymmetric flow is the density gradient, rather than the Mach number itself. In the highly asymmetric cases, the dense material from negative $y$ values does not stagnate at $y=0$, as in the canonical HL flow. Instead, this material pushes its way to positive $y$ values (where the background density is lower) as it is deflected by the gravitational influence of $M_2$. In the cases where  $\mathcal{M}_\infty = 1.15$, the flow is nearly symmetric as density gradients are quite mild and the flow morphology approaches that of the classic HL case.

The lower panels in Figures \ref{fig:sims_g43_fix_q_vary_mach} and \ref{fig:sims_g53_fix_q_vary_mach} show slices of flow Mach number near the embedded object. In case of the high upstream Mach numbers or steeper upstream density gradients, most of the material in the post-shock region is supersonic, with a negligible amount of material having $\mathcal{M} \ll 1$ values. As the upstream Mach number is decreased, or the upstream density made shallower, the bow shock becomes more symmetric. The upstream flow is supersonic, whereas after the material crosses the shock and meets the pressure gradient caused by the convergence of the flow in the post-shock region, the downstream flow becomes subsonic. In Figure \ref{fig:sims_g43_fix_q_vary_mach}, we observe that material re-crosses a sonic surface as it falls inward toward the sink; due to the difference in adiabatic index, this feature is not present in the $\gamma=5/3$ models of Figure \ref{fig:sims_g53_fix_q_vary_mach}.

We can anticipate the implications of these flow distributions on coefficients of accretion and drag. With increasing $\mathcal{M}_\infty$, the disturbance in the flow symmetry is expected to reduce the rate of accretion: streamlines show less material is converging toward the embedded object.  We also note that for larger density gradients (higher $\mathcal{M}_\infty$) the post-shock flow is generally more turbulent, and the rate of accretion of material into the sink becomes more variable. The variation of density flowing within the accretion radius in the high $\mathcal{M}_\infty$ cases lead dense material from negative $y$ regions to be focused into the object's wake, which might be expected to enhance the dynamical friction drag force.

\subsubsection{Dependence on Mass Ratio, $q_{\rm r}$}\label{sec:hydro_q}

\begin{figure*}
\centering
\includegraphics[width=16cm]{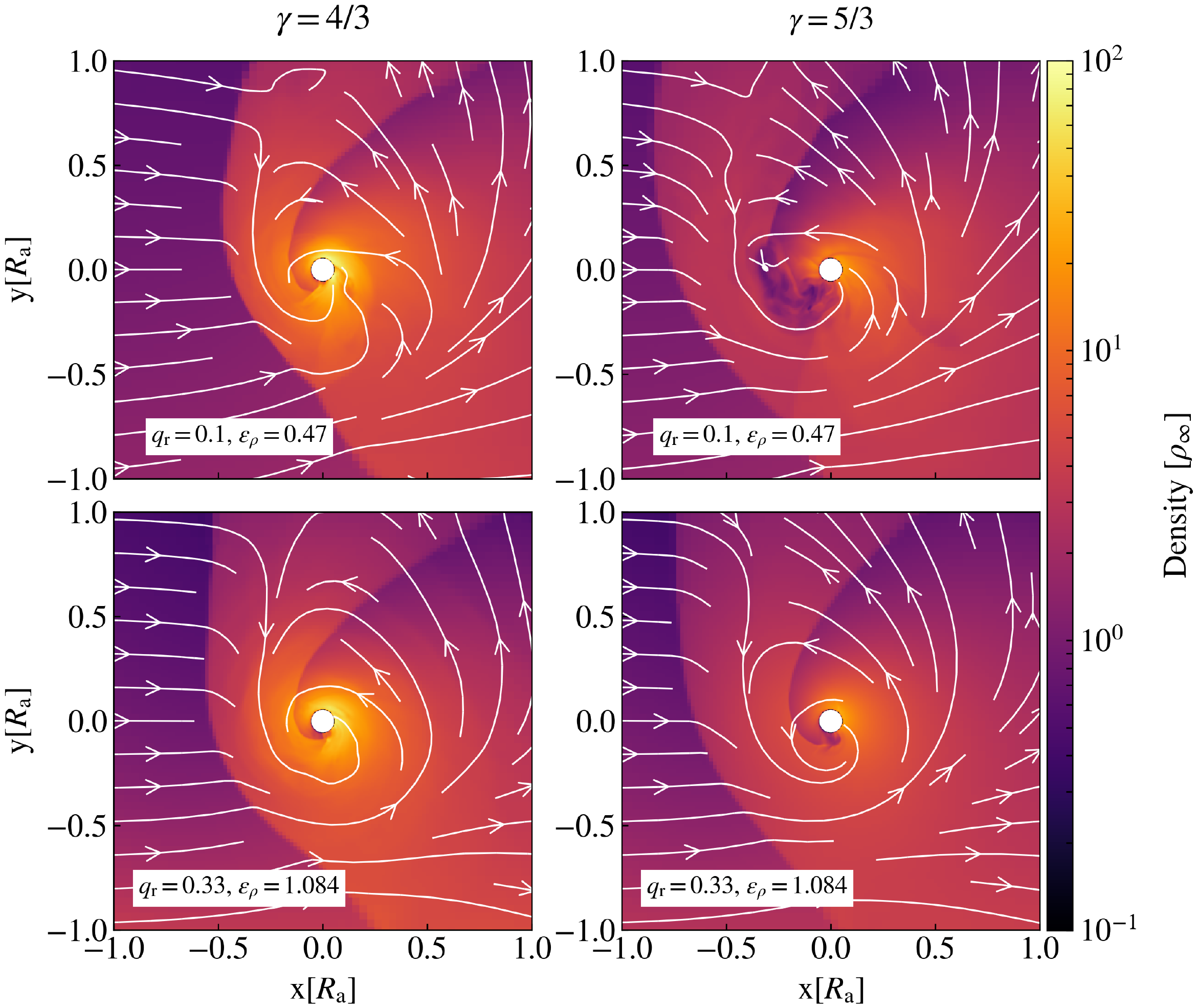}
\caption{Slices of density in units of $\rho_\infty$ for the simulation suite $(\Gamma_s, \gamma) = (4/3, 4/3)$ (left panels) and for the simulation suite $(\Gamma_s, \gamma) = (5/3, 5/3)$ (right panels) through the orbital ($x$-$y$) plane, for a fixed upstream Mach number $\mathcal{M}_\infty$ and varying ratio $q_{\rm r}$. The simulations for each $\gamma$ use $\mathcal{M}_\infty = 1.69$ and $q_{\rm r}=0.1$ (top) and 0.33 (bottom). The slices compare the state of the flow at simulation time $t = 30~R_{\mathrm{a}}/v_\infty$. The panels show the dependence of the flow pattern around the embedded companion object in a specific region of the envelope on the binary mass ratio. \label{fig:sims_fix_mach_vary_q}}
\vspace*{8mm}
\end{figure*}

Varying mass ratio can be representative of differing binary initial conditions, or even changing enclosed mass within a given binary.  
Figure~\ref{fig:sims_fix_mach_vary_q} shows slices of density  through the orbital ($x$-$y$) plane from the simulations performed for $q_{\rm r}$ values $1/10$ and $1/3$ and a fixed $\mathcal{M}_\infty = 1.69$ for both $\gamma=\Gamma_{\mathrm s}=4/3$ and $\gamma=\Gamma_{\mathrm s}=5/3$.

Comparison of the panels of Figure~\ref{fig:sims_fix_mach_vary_q} demonstrates the effect of $q_{\rm r}$ on the flow characteristics. 
Although $\mathcal{M}_\infty$ is held constant, the corresponding $\epsilon_\rho$ is largest in the $q_{\rm r} = 1/3$ case, and smallest for $q_{\rm r} = 1/10$, as shown in Tables \ref{tab:sims_43_params} and \ref{tab:sims_53_params}. This yields the most obvious difference with varying $q_{\rm r}$: the flow in the $q_{\rm r}=1/3$ case is more asymmetric (for example the bow shock is more distorted) as a result of the stronger density gradient. Secondly, we observe that the higher $q_{\rm r}$ cases have weaker focusing of the flow around the embedded object, as evidenced by the pre-shock flow streamlines. This happens because as the mass ratio increases, from equation~\eqref{eq:Ra_asfunc_a}, $R_{\mathrm a}/a$ increases. We choose our model domain sizes to capture this difference in scales, as described in Section \ref{sec:method}. When the accretion radius is a larger fraction of the orbit size, gravitational focusing acts over fewer characteristic lengths $R_{\rm a}$ to concentrate the flow. One implication is that the effective interaction cross section is smaller than $\pi R_{\mathrm{a}}^2$, because the derivation of $R_{\mathrm{a}}$ imagines a ballistic trajectory focused from infinite distance. 

Therefore, with increasing $q_{\rm r}$, we anticipate a decrease in the dynamical friction drag force due to the smaller effective cross section. The implications for the accreted mass are less obvious from these slices because the morphology of the post-shock flow is largely similar due to the competition between steeper density gradients but smaller effective cross sections at larger $q_{\rm r}$. 

\subsubsection{Dependence on Adiabatic Index, $\gamma$}\label{sec:hydro_gamma}

Here we examine the dependence of flow properties on the stellar envelope equation of state, using two limiting cases of ideal-gas equations of state that bracket the range of typical stellar envelope conditions. 
A $\gamma=4/3$ equation of state is representative of a radiation pressure dominated equation of state, occuring in massive-star envelopes, or in zones of partial ionization in lower-mass stars. A $\gamma=5/3$ equation of state represents a gas-pressure dominated equation of state, as occurs in the interiors of relatively low-mass stars with masses less than approximately $ 8 M_\odot$~\citep[e.g.][]{MacLeod:2017,Murguia-Berthier:2017}. Values between these limits are also possible, dependent on the microphysics of the density--temperature regime \citep{Murguia-Berthier:2017}.

While there are many similarities in overall flow morphology in our simulation suites A (Table \ref{tab:sims_43_params}) and B (Table \ref{tab:sims_53_params}), because gas is less compressible with $\gamma=5/3$ than it is with $\gamma=4/3$, there are a several key differences between these two cases.   Gas near the accretor tracks closer to ballistic, rotationally-supported trajectories in the $\gamma = 4/3$ case, as compared to the less compressible $\gamma = 5/3$ case. A related feature is that the bow shock stands further off from the accretor into the upstream flow when  $\gamma = 5/3$  than $\gamma=4/3$. These properties are visible when comparing the equivalent panels of Figure~\ref{fig:sims_g53_fix_q_vary_mach} and Figure~\ref{fig:sims_g43_fix_q_vary_mach}, or the left and right panels of Figure~\ref{fig:sims_fix_mach_vary_q}.  The underlying explanation is similar, shock structures around the accretor are set by the balance of the gravitational attraction of the accretor, the ram pressure of incoming material, and pressure gradients that arise as gas is gravitationally focused.  For the less-compressible $\gamma=5/3$ models, gas pressure gradients exceed the accretor's gravity, and partially prevent accretion. We observe the consequence of this in lower-density voids of hot, low Mach number material in Figure \ref{fig:sims_g53_fix_q_vary_mach}. For the more compressible $\gamma = 4/3$ flow, gas is more readily compressed, and pressure gradients build at a similar rate to the gravitational force \citep{Murguia-Berthier:2017}. One consequence of this is that higher densities near the accretor track the compression of gas deep into the accretor's gravitational potential well.

\vspace{2.5mm}
\subsection{Coefficients of Drag and Accretion}\label{sec:coeff}

\begin{figure*}
  \centering
  \includegraphics[width=\textwidth]{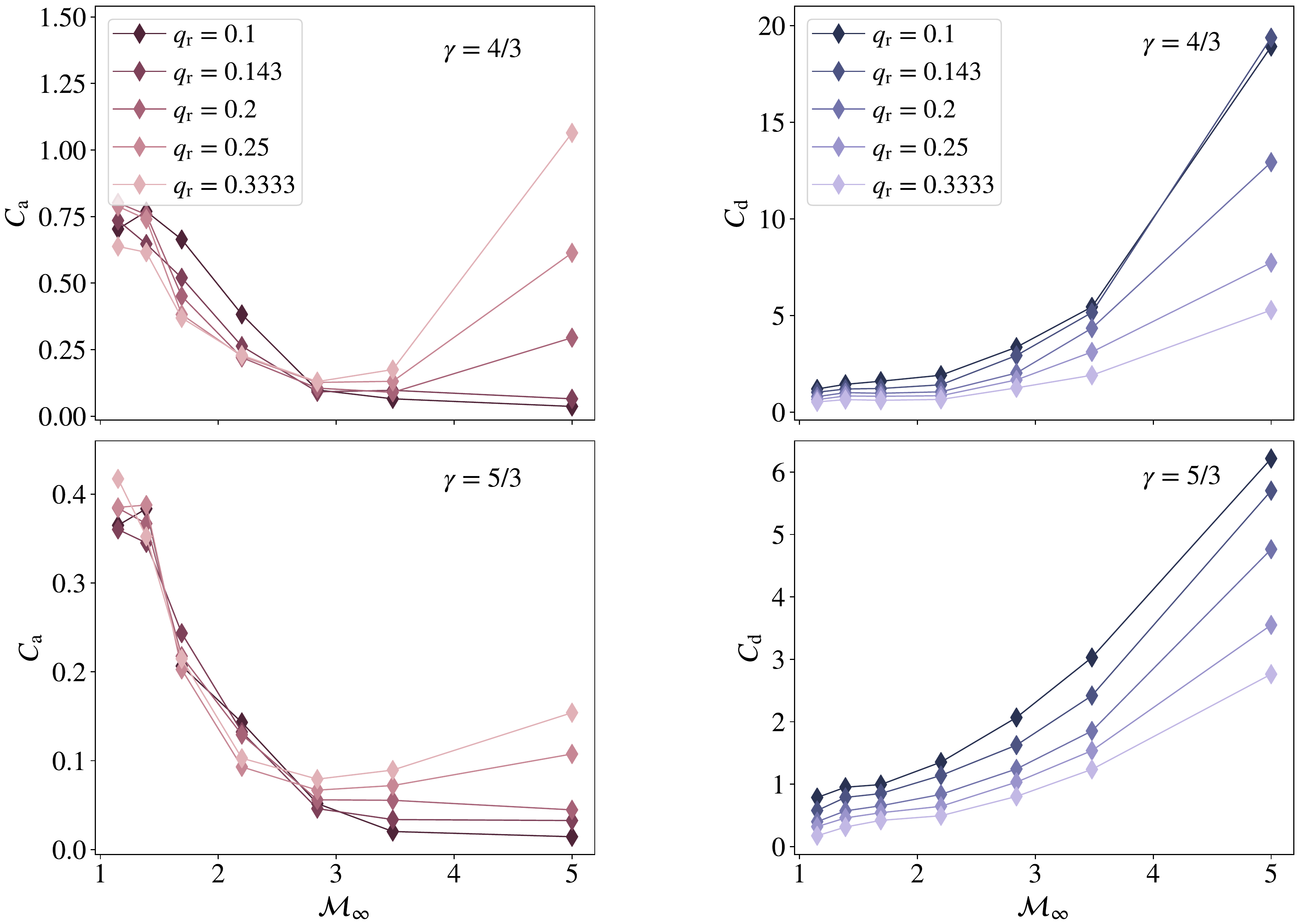}
  \caption{Variations of the median coefficient of accretion $C_{\mathrm a}$ and the median coefficient of drag $C_{\mathrm d}$ versus upstream Mach number $\mathcal{M}_\infty$ for the $(\Gamma_s, \gamma) = (4/3, 4/3)$ simulations (top panels) and $(\Gamma_s, \gamma) = (5/3, 5/3)$ simulations (bottom panels). $C_{\mathrm a} (C_{\mathrm d})$ vs. $~\mathcal{M}_\infty$ curves are shown for each $q_{\rm r}$ value at which simulations are performed. $C_{\mathrm a}$ is obtained by normalizing the mass accretion rate in the system to the HL theory mass accretion rate. $C_{\mathrm d}$ is obtained by normalizing the drag force in the system to the HL theory drag force. The $C_{\mathrm a}$ and $C_{\mathrm d}$ median values are computed in the simulation time range $10~R_{\mathrm{a}}/v_\infty < t < 30~R_{\mathrm{a}}/v_\infty$. For a fixed, small mass ratio $q_{\rm r}$, a higher $\mathcal{M_\infty}$ corresponds to a steeper upstream density gradient which breaks the symmetry of the flow, causing a reduction in $C_{\mathrm a}$; and a greater quantity of dense material gravitationally focused from the deep stellar interior, which increases $C_{\rm d}$. \label{fig:datapoints_Ca_Cd_vs_mach_g43_g53_sims}}
\vspace*{1cm}
\end{figure*}

We now use the results from the wind tunnel experiments to understand the effects of $q_{\rm r}$ and $\mathcal{M}_\infty$ on the accretion of material onto the embedded object and on the drag force acting on the embedded object. 
Figure~\ref{fig:datapoints_Ca_Cd_vs_mach_g43_g53_sims} shows median values of $C_{\mathrm a}$ and $C_{\mathrm d}$ computed over simulation times $10~R_{\mathrm{a}}/v_\infty < t < 30~R_{\mathrm{a}}/v_\infty$, as a function of $\mathcal{M}_\infty$ for different values of $q_{\rm r}$. We use contributions from both the dynamical friction drag force, $F_{\rm{df}}$, and the force due to linear momentum accretion $F_{\dot p_x}$ in calculating $C_{\rm d}$ (Equation \ref{eq:C_d}). In all our simulations, $F_{\rm{df}}$ is larger than $F_{\dot p_x}$, however, as we find in Section \ref{sec:sink}, the sum of these forces is the quantity that is invariant with respect to changing the numerical parameter of sink radius.

In Appendix \ref{subsec:fits}, we present fitting formulae for the coefficients of accretion $C_{\mathrm a}$ and drag force $C_{\mathrm d}$ as a function of the mass ratio and Mach number from both our $\gamma = 4/3$ and $\gamma = 5/3$ simulations, showing the mapping between the $(q_{\rm r}, \mathcal{M}_\infty) \rightarrow (C_{\mathrm a}, C_{\mathrm d})$ parameter space that we have explored.

\subsubsection{Dependence on Mach Number, ${\cal M}_\infty$}\label{sec:coeff_mach}

We begin by examining the dependence of drag and accretion coefficients with upstream Mach number, ${\cal M}_\infty$. 
Figure~\ref{fig:datapoints_Ca_Cd_vs_mach_g43_g53_sims} shows that for ${\cal M}_\infty \lesssim 3$, at fixed $q_{\rm r}$, $C_{\rm a}$ decreases with increasing $\mathcal{M}_\infty$. For $q_{\rm r}\lesssim0.2$, this trend continues to higher $\mathcal{M}_\infty$, while for $q_{\rm r} \gtrsim 0.2$, the coefficient of accretion rises again with increasing $\mathcal{M}_\infty$, particularly in the $\gamma=4/3$ models. This general trend can be understood in the context of the associated density gradients. For fixed $q_{\rm r}$, higher $\mathcal{M}_\infty$ flows correspond to steeper density gradients relative to the accretion radius. The steep density gradient breaks the symmetry of the post-shock flow, as discussed in Section \ref{sec:hydro_mach}. The resulting net rotation and  angular momentum act as a barrier to accretion, and lead to a drop in the accretion rate as compared to the HL rate ($\pi R_{\mathrm{a}}^2 \rho_\infty v_\infty$) \citep{MacLeod_2015}. The increase in $C_{\rm a}$ for large $q_{\rm r}$ at high $\mathcal{M}_\infty$  runs counter to this overall trend. In these cases, the combined steepening of the density gradient and weakening of the overall gravitational focus and slingshot discussed in Section \ref{sec:hydro_q} leads to a flow morphology that very effectively transports dense material from $-y$ impact parameters toward the sink, instead of imparting so much angular momentum that it is flung to $+y$ coordinates, resulting in large $C_{\rm a}$.

As for the drag force, we see that for each value of $q_{\rm r}$, $C_{\mathrm d}$ monotonically increases by a factor of $\mathcal{O}(10)$ with increasing $\mathcal{M}_\infty$ across the range of $\mathcal{M}_\infty$ values for which we have performed simulations. This trend reflects the fact that higher local gas densities, $\rho$, are achieved within the accretion radius of $M_2$ for higher values of the upstream Mach number, $\mathcal{M}_\infty$. This higher density material ($\rho \gg \rho_\infty$) focused into the wake of the embedded object from deeper inside the interior of the primary star enhances the dynamical friction drag force as compared to the HL drag force ($\pi R_{\mathrm{a}}^2 \rho_\infty v_\infty^2$).

\subsubsection{Dependence on Mass Ratio, $q_{\rm r}$}\label{sec:coeff_q}

For each $\mathcal{M}_\infty$, we can also see the dependence of $C_{\rm a}$ and $C_{\rm d}$ on the mass ratio $q_{\rm r}$ in Figure \ref{fig:datapoints_Ca_Cd_vs_mach_g43_g53_sims}. As the mass ratio increases, the accretion radius becomes a larger fraction of the orbit size. This causes the flows to be focused from a distance that is a smaller multiple of the accretion radius, causing weaker focusing and gravitational slingshot of the gas, as discussed in Section \ref{sec:hydro_q}. The effect of this difference on the coefficients of accretion at $\cal{M}_\infty \lesssim$ 3 is minimal. However, as discussed above in Section \ref{sec:coeff_mach},  at higher $\cal{M}_\infty$, there is a dramatic increase in $C_{\rm a}$ with increasing $q_{\rm r}$ that results in the capture of dense material from $-y$ impact parameters that does not possess sufficient momentum to escape the accretor's gravity. 

The counterpoint of weakened momentum transfer to the gas in the higher $q_{\rm r}$ cases is that the embedded object is impeded less by this gravitational interaction. In section \ref{sec:hydro_q}, we discussed this effect in terms of a reduced effective cross section. In terms of the coefficients of drag in Figure \ref{fig:datapoints_Ca_Cd_vs_mach_g43_g53_sims}, the quantitative effects are particularly clear. When gas is gravitationally focused over fewer characteristic length scales (because $R_{\rm a}$ is a larger fraction of $a$ at larger $q_{\rm r}$) we see lower dimensionless drag forces, $C_{\rm d}$. 

\subsubsection{Dependence on Adiabatic Index, $\gamma$}\label{sec:coeff_gamma}

The gas adiabatic index has important consequences for coefficients of drag and accretion because while pressure gradients enter into the fluid momentum equation, distributions of gas densities set rates of drag and accretion. Thus, the equation of state is crucial for both the flow morphology, as discussed in Section \ref{sec:hydro_gamma}, and for $C_{\rm a}$ and $C_{\rm d}$.

In Figure \ref{fig:datapoints_Ca_Cd_vs_mach_g43_g53_sims} we note that the increased resistance to compression by the accretor's gravitational force of the $\gamma = 5/3$ models leads to lower $C_{\rm a}$ by a factor of approximately 2 than the equivalent $\gamma=4/3$ models.   We saw the effects of this in the density slices of Figures \ref{fig:sims_g43_fix_q_vary_mach} and \ref{fig:sims_g53_fix_q_vary_mach}, in which the material in the vicinity of $M_2$ is not as dense in the $\gamma = 5/3$ models as it is in the $\gamma = 4/3$ simulations.  Second, the larger pressure support provided by the gas in the $\gamma = 5/3$ simulations decreases the overdensity of the post-shock wake versus what is realized in the simulations with $\gamma = 4/3$.  The greater upstream-downstream symmetry that results, decreases the net dynamical friction force exerted on the embedded object. We observe that $C_{\rm d}$ is approximately a factor of 3 lower for $\gamma=5/3$ than $\gamma=4/3$ in the right panels of Figure~\ref{fig:datapoints_Ca_Cd_vs_mach_g43_g53_sims}. 

Having explored the parameter space of gas flow and coefficients of gas and accretion in our wind tunnel models, in the following section, we explore the application of these results to astrophysical common envelope encounters. 

\vspace{5mm}
\section{Accretion onto Black Holes During a Common Envelope Inspiral}\label{sec:implications}
In this section we discuss the application of our wind tunnel results to the scenario of a black hole dynamically inspiraling through the envelope of its companion. We focus in particular on the accreted mass and spin, because these parameters directly enter into the gravitational-wave observables.  To do so, we discuss the application and extrapolation of our numerical measurements of $C_{\rm a}$ and $C_{\rm d}$ to black holes, and the implications on the accreted mass and spin for LIGO-Virgo's growing binary black hole merger population. 

\vspace*{2.5mm}
\subsection{Projected Accretion and Drag Coefficients for Compact Objects}\label{sec:sink}

A limitation of our numerical models is that the accretion rate, and to a lesser extent the drag force, have been shown to depend on the size of the central absorbing sink \citep[see][]{1994ApJ...427..351R, 1994A&AS..106..505R,1995A&AS..113..133R, 2012ApJ...752...30B,MacLeod_2015,Antoni:2019pgq}. This dependence indicates that results do not converge to a single value regardless of the numerical choice of sink radius, $R_{\rm s}$. Further, simultaneously resolving the gravitational focusing radius, $R_{\rm a}$, and the size of a compact object is currently not computationally feasible: $R_{\rm a}$ might be on the order of the envelope radius, while an embedded compact object's radius orders of magnitude smaller still. Previous work by \citet{MacLeod:2014yda,MacLeod_2015} and \citet{MacLeod:2017} has pointed out that these limitations make accretion coefficients derived from simulations at most upper limits on the realistic accretion rate. 

Here we attempt to systematically explore the scaling of coefficients of accretion and drag to smaller sink radii, that is smaller $R_{\rm s}/R_{\rm a}$. We ran two additional sets of 35 models that reproduce models A1 through A35, reducing the sink radius by a factor of two to $R_{\rm s}/R_{\rm a}=0.025$ and $R_{\rm s}/R_{\rm a}=0.0125$. To preserve the same level of resolution across the sink radius, we add an additional layer of mesh refinement around the sink with each reduction of sink radius (effectively halving the minimum zone width). From these models, we measure coefficients of drag and accretion following the methodology identical to our standard models presented earlier. 

With accretion and drag coefficients derived across a factor of four in sink radius, we fit the dependence on sink radius with power laws of the form 
\begin{equation}\label{eq:powerlawacc}
    \log_{\rm 10}\left( \dot M \right) = {\alpha_{\dot M}} \log_{\rm 10}\left( R_{\rm s}/R_{\rm a} \right) + {\beta_{\dot M}}
\end{equation}
\begin{equation}\label{eq:powerlawdrag}
    \log_{\rm 10}\left( F_{\rm d} \right) = {\alpha_{\rm F}} \log_{\rm 10}\left( R_{\rm s}/R_{\rm a} \right) + {\beta_{\rm F}}.
\end{equation}
Thus $\dot M \propto \left(R_{\rm s}/R_{\rm a}\right)^{\alpha_{\dot M}}$ and $F_{\rm d} \propto \left(R_{\rm s}/R_{\rm a}\right)^{\alpha_{\rm F}}$. With these coefficients, we have some indication of how rates of accretion and drag forces might extrapolate to much smaller $R_{\rm s}/R_{\rm a}$ that are astrophysically realistic. 

\begin{figure}[tbp]
 \includegraphics[width=0.49\textwidth]{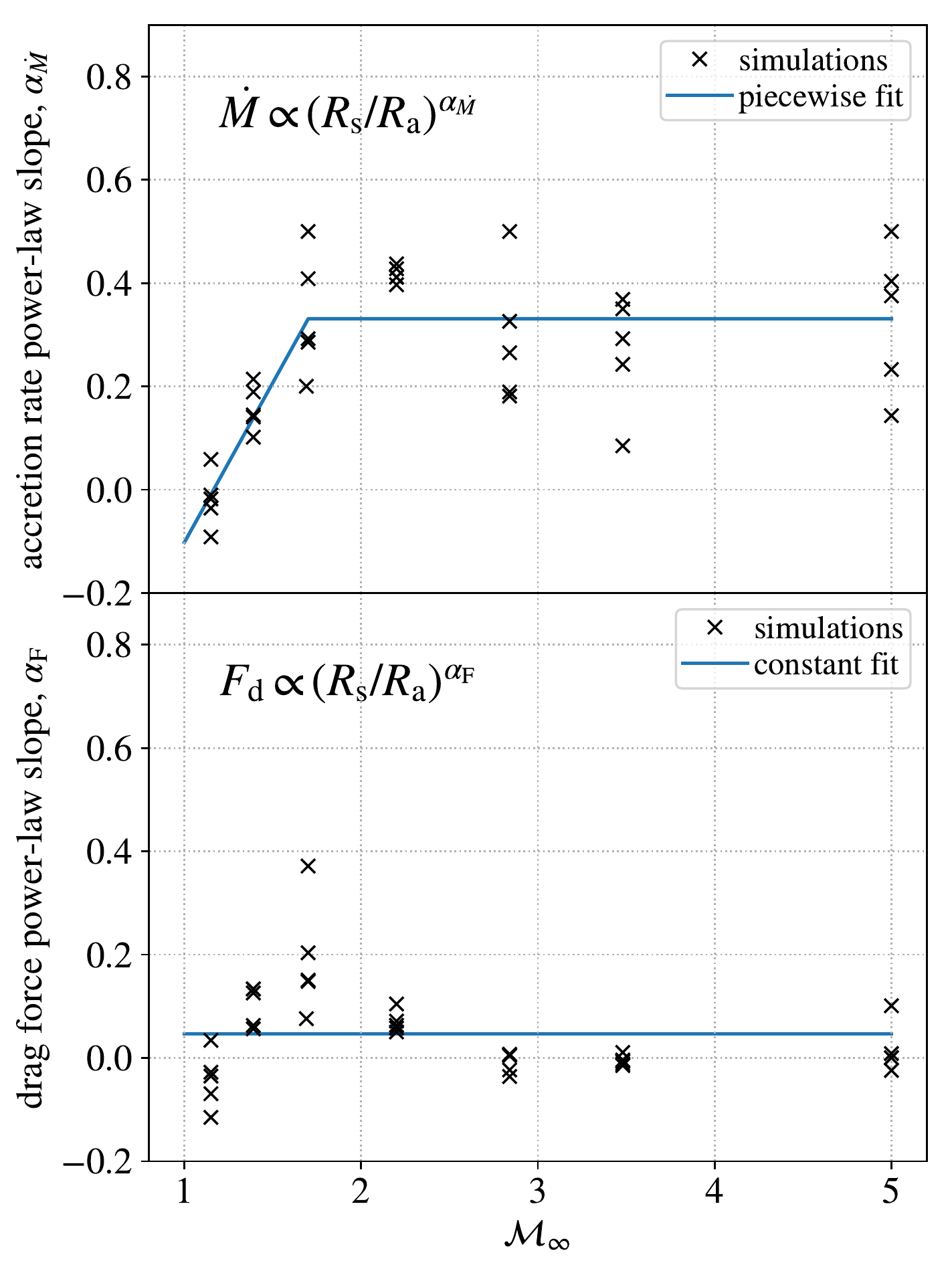}
 \caption{The figure shows exponents of power-law relations of the accretion rate, $\alpha_{\dot M}$, and drag force, $\alpha_{\rm F}$, on the sink radius. Top panel: For each $\gamma = 4/3$ simulation in Table~\ref{tab:sims_43_params}, an $\alpha_{\dot M}$ (denoted by a cross marker) is derived from a linear least-squares fit of Eq.~\eqref{eq:powerlawacc} to the $C_{\rm a}$ value measured from that simulation of sink size $R_{\rm s} = 0.05~R_{\rm a}$, plus similar simulations with $R_{\rm s} = [0.025, 0.0125]~R_{\rm a}$. The plot shows a largely-positive $\alpha_{\dot M}$, which indicates that accretion rates decrease as sink size decreases. The blue line shows a fit to the dependence of $\alpha_{\dot M}$ on $\cal {M}_\infty$, using the piecewise fitting relation given by Eq.~\eqref{eq:piecewise}.
 Lower panel: Following the same procedure for calculating $\alpha_{\dot M}$, each $\alpha_{\rm F}$ (denoted by a cross marker) value is derived from a linear least-squares fit of Eq.~\eqref{eq:powerlawdrag} to $C_{\rm d}$ values from simulations of varying sink sizes. The plot shows values of $\alpha_{\rm F} \sim 0$, indicating little change in the overall drag force as the sink radius is modified. The blue line shows $\alpha_{\rm F}$ across $\cal {M}_\infty$ values using a constant least-squares fit, which gives $\alpha_{\rm F} \approx 0.05$.
 \label{fig:sink} }
\vspace{5mm}
\end{figure}

Figure \ref{fig:sink} presents the exponents of the power-law relations of the accretion rate and drag force on the sink radius, as a function of ${\cal M}_{\infty}$.  For each ($q_{\rm r}, \cal M_\infty$) model, there are three sets of ($C_{\rm a}, C_{\rm d}$) values from the $R_{\rm s}/R_{\rm a}$ = [0.0125, 0.025, 0.05] simulations respectively. A linear least-square fit of Eq.~\ref{eq:powerlawacc} to the three $C_{\rm a}$ values is performed. The slope of the fitted line is $\alpha_{\dot M}$, that is the exponent of the power law function relating $\dot M$ to $R_{\rm s}/R_{\rm a}$. Similarly, a linear least-square fit of Eq.~\ref{eq:powerlawdrag} to the three $C_{\rm d}$ values is used to derive $\alpha_{\rm F}$, the exponent of the power law function relating $F_{\rm d}$ to $R_{\rm s}/R_{\rm a}$. Thus, we derive one $\alpha_{\dot M}$ and one $\alpha_{\rm F}$ (represented with cross markers in Fig.~\ref{fig:sink}) per ($q_{\rm r}, \cal{M}_\infty$) model. We observe that the majority of the $\alpha_{\dot M}$ values are positive, indicating that accretion rates drop as sink sizes get smaller relative to $R_{\rm a}$. Additionally, we observe that $\alpha_{\dot M}$ is typically lower in low Mach number flows, ${\cal M}_\infty \lesssim 2$, which have proportionately shallower density gradients. Above  ${\cal M}_\infty \gtrsim 2$, $\alpha_{\dot M}$ is approximately constant with increasing ${\cal M}_\infty$. At a given ${\cal M}_\infty$, there is variation between the models, depending on the mass ratio, $q_{\rm r}$. However, for simplicity, the following piecewise linear plus constant least-squares fit (blue line in Figure \ref{fig:sink}) reproduces the main trends
\begin{equation}\label{eq:piecewise}
\alpha_{\dot M} \approx 
\begin{cases}
0.62 {\cal M}_\infty - 0.72,& \ \ {\cal M}_\infty < 1.7, \\
0.33,& \ \ {\cal M}_\infty \geq 1.7 .
\end{cases}
\end{equation}

By comparison, exponents of power-law dependence of the drag coefficients on sink radius, $\alpha_{\rm F}$, do not show particularly structured behavior with ${\cal M}_\infty$. Further, most values are near zero, with all but one model lying within $-0.2 < \alpha_{\rm F} < 0.2$. Least-squares fitting of a constant finds $\alpha_{\rm F} \approx 0.05$, that is close to 0. This indicates that there is little change in the drag force with changing sink size.

Taken together, these scalings indicate that when $R_{\rm s}/R_{\rm a} \ll 1$, we can expect drag forces to remain relatively unchanged while accretion rate decreases. As a specific example, if an accreting black hole has $R_{\rm s}/R_{\rm a}=10^{-5}$ at ${\cal M}_\infty=2$, our scaling above suggests that we can expect the realistic accretion coefficient to be approximately 6\% of the value derived in our simulations with $R_{\rm s}/R_{\rm a}=0.05$ (because $(10^{-5}/0.05)^{0.33}\approx 0.06$).  This result makes intuitive sense in light of our simulation results: drag forces arise from the overdensity on the scale of $R_{\rm a}$, while, especially in the higher ${\cal M}_{\infty}$ (higher $\epsilon_\rho$) cases, rotation inhibits radial, supersonic infall of gas to the smallest scales.  

\vspace*{2.5mm}
\subsection{Coupled Orbital Tightening and Accretion}\label{sec:coupled}
As a black hole spirals through the common envelope gas, its orbit tightens in response to drag forces, and it may also potentially accrete mass from its surroundings. Under the HL theory of mass accretion and drag, the degree of mass growth is coupled to the degree of orbital tightening. Thus, a given orbital transformation is always accompanied by a corresponding mass change in this theory. \citet{Chevalier:1993,Brown:1995,Bethe:1998} elaborated on this argument, and suggested that compact objects in common envelope phases might easily double their masses. 

Here we re-express this line of argument with the addition of separate coefficients of drag and accretion (which might, for example, be motivated by numerical simulations). Orbital energy, $E = -GM_1 M_2 / 2 a$, is dissipated by the drag force at a rate $\dot E = - F v$ (if force is defined positive, as in our notation). Expressed in terms of the coefficient of drag, $\dot E = - C_{\rm d} F_{\rm HL} v = - C_{\rm d} \dot M_{\rm HL} v^2 = - C_{\rm d} \dot E_{\rm HL}$ (equations \eqref{eq:mdotHL} and \eqref{eq:fHL}). We will approximate the relative velocity here as the Keplerian velocity, such that $v^2 \approx G(M_1 + M_2)/a$. We can then write the mass gain per unit orbital energy change,
\begin{align}
\frac{dM_2}{dE} &=\frac{\dot M}{\dot E} = - \frac{C_{\rm a} \dot M_{\rm HL} }{C_{\rm d} \dot M_{\rm HL} v^2} = - \frac{C_{\rm a}  }{C_{\rm d}  v^2}, \nonumber \\
& = \frac{1}{2(1+q_{\rm r})} \frac{M_2}{E} \frac{C_{\rm a}  }{C_{\rm d} },
\end{align}
or equivalently, 
\begin{equation}\label{eq:dlnMdE}
 \frac{d \ln M_2}{d \ln E}   = \frac{1}{2(1+q_{\rm r})} \frac{C_{\rm a}  }{C_{\rm d} }.
\end{equation}
This implies that the mass gained by the embedded, accreting compact object is related to the reduced mass of the pair, and the ratio of accretion to drag coefficients. 
We can integrate this equation under the approximation that $q_{\rm r}$, $C_{\rm a}$ and $C_{\rm d}$ remain close to typical values, which we denote $\overline{C_{\rm a}}$, $\overline{C_{\rm d}}$ and $\overline{q_{\rm r}}$, over the course of the inspiral from the onset of common-envelope evolution through envelope ejection.  In this approximation,
\begin{equation}\label{eq:MEsolution}
\frac{M_{2,f}}{M_{2,i}} \approx \left(\frac{E_f}{E_i}\right)^ {\left(\frac{1}{2(1 +\overline{q_{\rm r}})} \frac{\overline{C_{\rm a}}}{\overline{C_{\rm d}}}\right)}.
\end{equation}
We can therefore conclude that if $\overline{C_{\rm a}}=\overline{C_{\rm d}}=1$, the fractional change in the mass of the embedded object is on the order of the square root of the change in the orbital energy, i.e., binary separation~\citep{Chevalier:1993,Brown:1995,Bethe:1998}. 

If accreted material carries net angular momentum, a black hole will also accrue spin. Assuming an initially non-spinning black hole, the accrued spin can be written in terms of $\Delta M_2 / M_{2,i}$. The highest spins are achieved if material accretes with the specific angular momentum of the last stable circular orbit and uniform direction. In this case, the final spin is 
\begin{equation}\label{eq:aspin}
    \chi = \sqrt{2 \over 3} X \left( 4 - \sqrt{18 X^2 - 2} \right),
\end{equation}
where $X=1/(1 + \Delta M_2 / M_{2,i})$  \citep{kingkolb:1999}.
Under these assumptions, the dimensionless spin reaches unity when $X = 1/\sqrt{6}$ or $\Delta M / M_{2,i}\approx 1.4$ \citep[as shown in Figure 1 of][]{kingkolb:1999}. 

From these arguments, we see that the ratio of accretion to drag coefficient is crucial in determining the accrued mass and spin onto a compact object. In the HL formalism, in which $\overline{C_{\rm a}}=\overline{C_{\rm d}}=1$, and is the accreted mass is given by equation \eqref{eq:MEsolution}, for $\overline{q_{\rm r}}=0.1$, we find that $\chi\rightarrow1$ for $E_f/E_i \gtrsim 7$.

\subsection{Implications for CE-transformation of Black Holes and Gravitational-Wave Observables}\label{sec:LIGO}

\begin{figure}[t]
  \includegraphics[width=\columnwidth]{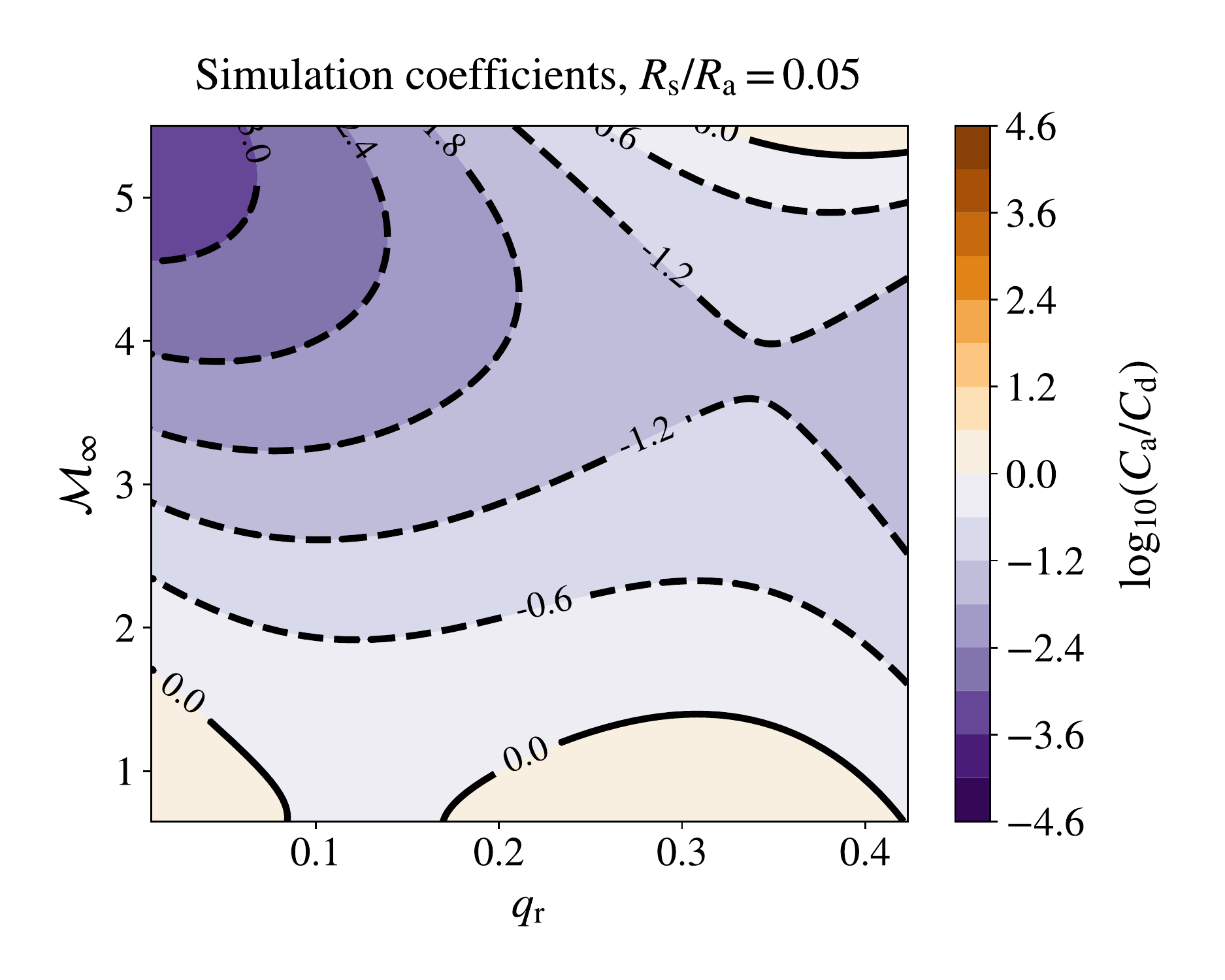}
  \includegraphics[width=\columnwidth]{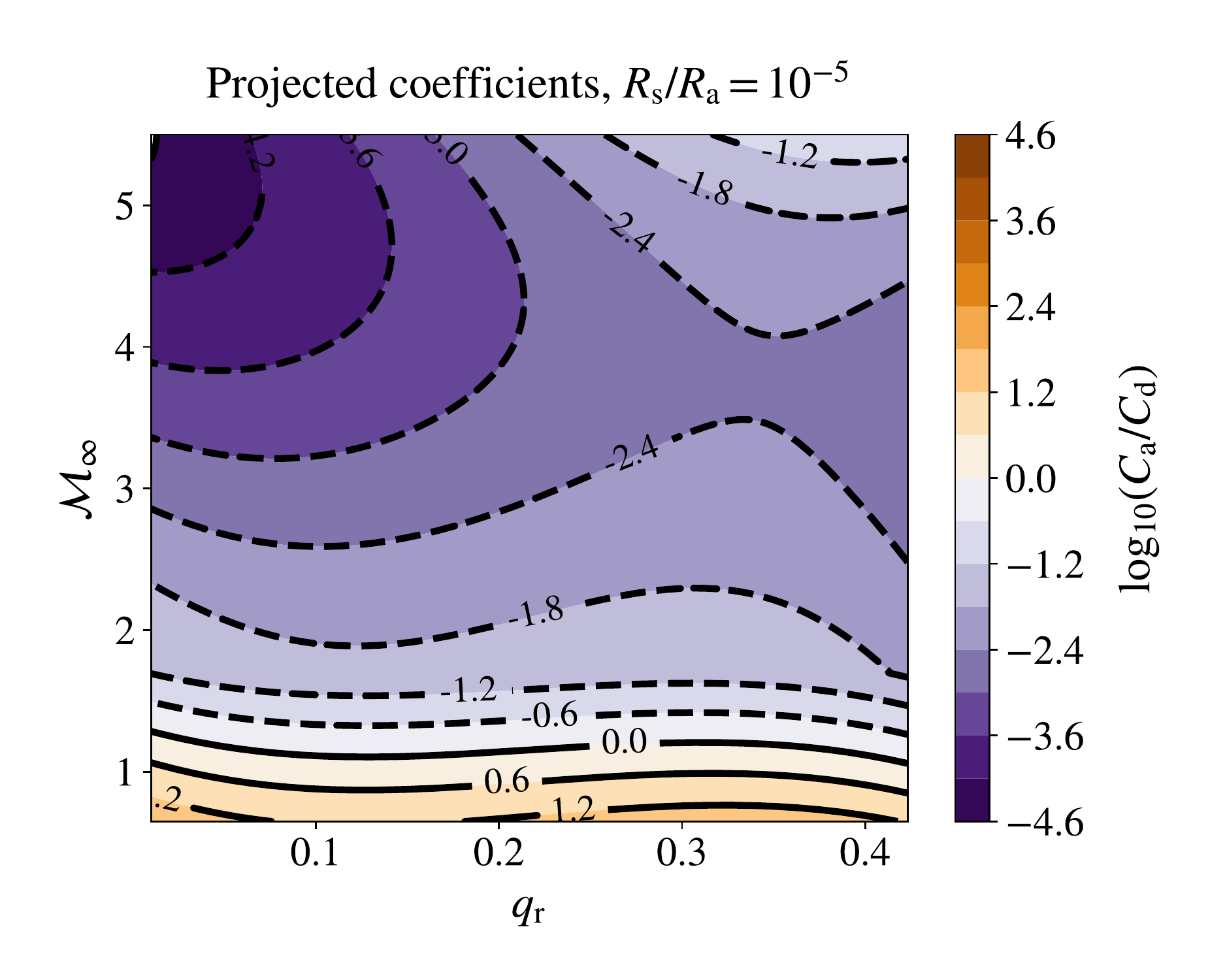}
\caption{Two-dimensional contour plot of $\log_{10}(C_{\mathrm a}/C_{\mathrm d})$ in the $q_{\rm r}-\mathcal{M}_\infty$ space using numerical results from the $\gamma = 4/3$ simulations presented in this paper. The top panel shows the $(q_{\rm r}, \mathcal{M}_\infty) \rightarrow (C_{\mathrm a}, C_{\mathrm d})$ mapping for the sink size used in the simulations $R_{\rm s}/R_{\rm a} = 0.05$. The bottom panel shows the $(q_{\rm r}, \mathcal{M}_\infty) \rightarrow (C_{\mathrm a}, C_{\mathrm d})$ mapping with the coefficients extrapolated to a sink size $R_{\rm s}/R_{\rm a} = 10^{-5}$, which is more realistic for a black hole embedded in a common envelope.\label{fig:mdot_fdf_ratio_rs05_rs1e-5}}
\vspace{5mm}
\end{figure}

In Figure \ref{fig:mdot_fdf_ratio_rs05_rs1e-5} we show the ratio of the coefficients of drag and accretion derived in our simulations. For illustrative purposes, we also scale these values using the power-law slopes derived in Section \ref{sec:sink} to a much smaller sink radius, $R_{\rm s}/R_{\rm a}= 10^{-5}$. This is, for example, appropriate for a $5M_\odot$ black hole (with horizon radius of approximately $1.5\times10^6$~cm) embedded deep within a $30 M_\odot$ primary-star envelope at a separation of a $10R_\odot$. Then $q_{\rm r} = 1/6$, and $R_{\rm a}/a\approx 0.3$, from equation \eqref{eq:Ra_asfunc_a}. Thus $R_{\rm a} \approx 2 \times 10^{11}$~cm, and $R_{\rm s}/R_{\rm a} \sim 10^{-5}$. However, we note that for larger separations, even smaller $R_{\rm s}/R_{\rm a}$ will be appropriate.

We observe that for the majority of the $q_{\rm r}-{\cal M}_\infty$ parameter space, $C_{\rm a}/C_{\rm d} \ll 1$, even in the direct simulation coefficients, though $C_{\rm a}/C_{\rm d}$ approaches unity as ${\cal M}_\infty\rightarrow1$. For specificity, if we use our direct (unscaled) simulation coefficients, and take the example case of a black hole involved in a $\overline{q_{\rm r}}\sim 0.1$ encounter, $C_{\rm a}/C_{\rm d} \lesssim 0.1$ for ${\cal M}_\infty \gtrsim 2$.  This is the bulk of the relevant parameter space for a dynamical inspiral if the relative velocity between the black hole and the envelope gas is similar to the Keplerian velocity (see Figure 3 and 4 of \citet{MacLeod_2015} and Figure 1 of \citet{MacLeod:2017}). If $C_{\rm a}/C_{\rm d}=0.1$, then $d\ln M_2 / d\ln E \approx 0.045$ (equation \eqref{eq:dlnMdE}), i.e., a 5\% change in the mass of the black hole due to accretion per orbital e-folding during the common envelope encounter. If this accreted mass is coherently maximally rotating, the black hole would spin up to $\chi\sim 0.15$ (if it begins with $\chi=0$). Thus, if the orbital energy changes by a factor of 25, the black hole would accrete about 15\% of its original mass (equation \eqref{eq:MEsolution}) and spin up to $\chi\sim 0.4$ (equation \eqref{eq:aspin}).

However, we have argued in Section \ref{sec:sink} that the simulated $C_{\rm a}/C_{\rm d}$ can be misleadingly high (or, alternatively is best interpreted as a strict upper limit) because the compact object radius is orders of magnitude smaller than $R_{\rm s}$. With the rescaled results of the second panel of Figure \ref{fig:mdot_fdf_ratio_rs05_rs1e-5} for $R_{\rm s} = 10^{-5} R_{\rm a}$, we see that for the same $\overline{q_{\rm r}}=0.1$ encounter in which ${\cal M}_\infty \gtrsim 2$, the ratio of the accretion to drag coefficient is $C_{\rm a}/C_{\rm d} \lesssim 10^{-2}$. This in turn implies that a black hole undergoing such a common envelope encounter accretes according to $d\ln M_2 / d\ln E \approx 0.0045$. Again, taking the example of orbital energy changing by a factor of 25, the black hole would accrete 1.4\% of its own mass and spin up to $\chi \sim 0.05$.  Even if the binary hardens by three orders of magnitude during the common-envelope phase, a non-spinning black hole would only accrete $\sim 3\%$ of its original mass and spin up to $\chi \sim 0.1$.

A possible exception to these predictions of low accreted mass and spin are black holes embedded in ${\cal M}_\infty \sim 1$ flows (involving dense stellar envelope material) and proportionately shallow density gradients. In these cases black holes can accrete at similar to the HL rate, largely because the environment is nearly homogeneous on the scale of $R_{\rm a}$. This regime of Mach numbers may be relevant to the self-regulated common-envelope inspiral phase that follows the dynamical inspiral. However, in this case, Mach numbers are lower in part because the embedded objects interact with  much lower density, higher entropy gas as the orbit starts to stabilize~\citep[e.g.][]{Ohlmann:2016b,Ivanova:2016, 2018MNRAS.477.2349I,Chamandy:2019psk}. This is presented quantitatively in \citet{Chamandy:2019psk}'s study of forces during a common envelope simulation, which showed that forces significantly decrease below those expected from the original stellar profile as the orbit stabilizes. 

The current catalog of gravitational-wave events observed by the LIGO-Virgo detectors demonstrates the existence of moderately massive black holes in binary systems~\citep{Abbott:2016blz,TheLIGOScientific:2016pea,Nitz:2018imz,Biwer:2018osg,Abbott:2017vtc,Abbott:2017gyy,Abbott:2017oio,Zackay:2019tzo,Venumadhav:2019lyq,LIGOScientific:2018mvr,De:2018zrk}. Common-envelope evolution is considered to be one of the preferred channels for the formation of these binaries~\citep{Kruckow:2016tti,Belczynski:2016obo,EldridgeStanway:2016,Stevenson:2017,Mapelli:2018}. These predictions therefore have important potential implications when considering the evolutionary history of the LIGO-Virgo network's growing population of gravitational-wave merger detections. 

If the typical black hole passing through a common envelope-phase accreted a significant fraction of its own mass, and reached dimensionless spin near unity (as implied by equations \eqref{eq:MEsolution} and \eqref{eq:aspin} if $C_{\rm a}/C_{\rm d}=1$) this would have two directly observable consequences on the demographics of merging black holes. The mass gaps believed to exist in the birth distributions of black holes masses~\citep{Bailyn:1998,Kreidberg:2012,Ozel:2010,Farr:2011,Yusof:2013,Belczynski:2014,Marchant:2016,Woosley:2017} would be efficiently eradicated if black holes doubled their masses over the typical evolutionary cycle. Secondly, the average projected spins of merging black holes onto the orbital angular momentum would be large ($\chi_{\rm eff}\sim 1$ if coherently oriented) or at least broadly-distributed (if randomly oriented), contrary to the existing interpretation of spins from LIGO--Virgo black hole observations \citep[e.g.,][]{Farr:2017uvj,Farr:2017gtv,Tiwari:2018qch,Piran_2019}, or the predictions of spins in merging binary black holes~\citep[e.g.,][]{Bavera:2019fkg,Fuller:2019sxi,Zaldarriaga:2017qkw,Kushnir:2016zee,Schroder:2018hxk,Batta:2019clm}. 

Our prediction of percent-level mass and spin accumulation yields a very different landscape of post common-envelope black holes. Our models suggest that common envelope phases should not significantly modify the natal masses or spins of black holes. If black holes are formed with non-smooth  mass distributions (including gaps or other features) or with low spin values, our models predict that these features would persist through a common envelope phase. 

\vspace{5mm}
\section{Conclusions}\label{sec:conclusions}
In this paper have we explored the effects of varying the binary mass ratio on common envelope flow characteristics, as well as coefficients of accretion and drag, using the Common Envelope Wind Tunnel setup of \citet{MacLeod:2017}. As the binary mass ratio is varied, the ratio of the gravitational focusing scale of the flow to the binary separation changes.
We have also varied the flow upstream Mach number and gas adiabatic constant, which were investigated in \citet{MacLeod:2017} and \citet{MacLeod_2015}. We have derived fitting formulae for the efficiency of accretion and drag from our simulations, and have applied these to derive implications for the mass and spin accreted by black holes during the common envelope encounter. Some key conclusions of this work are:

\begin{enumerate}
\setlength\itemsep{1em}
\item Using a systematic survey of the dimensionless parameters that characterize gas flows past objects embedded within common envelopes, we use our simplified Common Envelope Wind Tunnel hydrodynamic model to study the role of the upstream Mach number $\mathcal{M}_\infty$, enclosed mass ratio $q_{\rm r}$, and the equation of state (as bracketed by adiabatic indices $\gamma=4/3$ and $\gamma=5/3$). For each model, we derive time-averaged coefficients of accretion, $C_{\rm a}$, and drag, $C_{\rm d}$ (Tables \ref{tab:sims_43_params} and \ref{tab:sims_53_params}).

\item Upstream Mach number $\mathcal{M}_\infty$ is a proxy for the dimensionless upstream density gradient $\epsilon_\rho$ (equation \ref{eq:mach-erho}). Higher $\mathcal{M}_\infty$ flows tend to have more asymmetric geometries due to steeper density gradients (Figures \ref{fig:sims_g43_fix_q_vary_mach} and \ref{fig:sims_g53_fix_q_vary_mach}). This transition in flow morphology is accompanied by higher drag coefficients but lower accretion coefficients (Figure \ref{fig:datapoints_Ca_Cd_vs_mach_g43_g53_sims}). 

\item The gas equation of state, parameterized here by the adiabatic index of ideal-gas hydrodynamic models $\gamma$, primarily affects the concentration of gas flow around the accretor. When $\gamma = 5/3$, pressure gradients partially act against gravitational focusing (Figure \ref{fig:sims_g53_fix_q_vary_mach} as compared to \ref{fig:sims_g43_fix_q_vary_mach}) and reduce coefficients of both accretion and drag by a factor of a few relative to $\gamma=4/3$ (Figure \ref{fig:datapoints_Ca_Cd_vs_mach_g43_g53_sims}). 

\item The binary mass ratio affects the ratio of gravitational focusing length to binary separation, $R_{\rm a}/a$, shown in equation \eqref{eq:Ra_asfunc_a} and Figure \ref{fig:ra-q}. As a result, larger mass ratio cases have weaker focusing of the flow around the embedded object, because gravitational focusing acts over a smaller number of gravitational focusing lengths to concentrate the flow (Figure \ref{fig:sims_fix_mach_vary_q}). The consequences of this distinction are reduced drag (lower $C_{\rm d}$) because of reduced momentum exchange with the flow, and, especially in the highest $\mathcal{M}_\infty$ cases, higher capture fractions (increased $C_{\rm a}$) because gas does not receive a sufficient gravitational slingshot to escape the accretor (Figure \ref{fig:datapoints_Ca_Cd_vs_mach_g43_g53_sims}).

\item The size of a typical accretor is a factor of $10^3$ to $10^8$ times smaller than the gravitational focusing radius, $R_{\rm a}$. Due to the limits of computational feasibility, our default numerical models adopt $R_{\rm s}/ R_{\rm a}=0.05$. We re-run the $\gamma=4/3$ models with $R_{\rm s}/ R_{\rm a}=0.025$ and $R_{\rm s}/ R_{\rm a}=0.0125$. We find that drag coefficients are insensitive to $R_{\rm s}$, but accretion coefficients have a dependence which we parameterize with a power-law slope, $\alpha_{\dot M}$ (Figure \ref{fig:sink}).
These scalings allow us to extend our Common Envelope Wind Tunnel results to more astrophysically realistic scenarios.

\item The amount of mass accreted by a compact object during a common envelope phase is coupled to the degree of orbital tightening, as per the HL theory \citep[and Section \ref{sec:coupled}]{Chevalier:1993,Brown:1995,Bethe:1998}. Angular momentum carried by the accreted mass may also spin up the object. Therefore, the values of $C_{\rm a}$ and $C_{\rm d}$ are crucial in determining the mass and spin accrued by embedded objects during the common envelope phase (specifically, the ratio $C_{\rm a}/C_{\rm d}$ sets the mass gain per unit orbital tightening, equation \eqref{eq:dlnMdE}). In the Hoyle Lyttton scenario, where $C_{\rm a}/C_{\rm d}=1$, the typical black hole immersed in a common envelope would gain on the order of its own mass and spin up to $\chi=1$.

\item Our simulation results that $C_{\rm a}/C_{\rm d} \ll 1$ suggest that black holes spiralling in through common envelopes accumulate less than 1\% mass per logarithmic change in orbital energy. In a typical event, this might correspond to a 1--2\% growth in black-hole mass and spin-up to a dimensionless spin of $\sim$ 0.05 for an initially non-spinning black hole (Figure \ref{fig:mdot_fdf_ratio_rs05_rs1e-5} and Section \ref{sec:LIGO}). Thus, our predictions suggest that common-envelope phases should not modify the mass and spin distributions of black holes from their natal properties.
\vspace{0.5cm}
\end{enumerate}

The hydrodynamic models presented in this paper have numerous simplifications relative to the complex, time-dependent geometry and flow likely realized in a common envelope interaction. Nonetheless, they allow us to discover trends by systematically exploring the parameter space that may arise in typical interactions. A companion paper, \citet{Rosa:2019}, considers the stellar evolutionary conditions for donor stars in common envelope systems under which this dimensionless treatment is useful. 

The ratio of accretion to drag coefficients (relative to their HL values) determines the amount of mass accretion during the dynamical inspiral phase of common envelope evolution. If our finding that $C_{\rm a}/C_{\rm d} \ll 1$ is correct, then the implications of this for gravitational wave observables are significant. In particular, if the birth mass distributions of black holes have non-smooth features, including gaps, or if black holes have low natal spins, these characteristic distributions will be preserved after the common envelope phase. 

\acknowledgements{We gratefully acknowledge helpful discussions with A. Murgia-Berthier, P. Macias, A. Frank, E. Blackman, and D. Brown. We thank the Niels Bohr Institute for its hospitality while part of this work was completed, and acknowledge the Kavli Foundation and the DNRF for supporting the 2017 Kavli Summer Program. S.D. received support for this work by the U.S. National Science Foundation grant PHY-1707954, the Inaugural Kathy '73 and Stan '72 Walters Endowed Fund for Science Research Graduate Fellowship, and the Research Excellence Doctoral Fellowship at Syracuse University. S.D. also thanks the Kavli Institute for Theoretical Physics (KITP) where portions of this work were completed. KITP is supported in part by the National Science Foundation under Grant No. NSF PHY-1748958. M.M. is grateful for support for this work provided by NASA through Einstein Postdoctoral Fellowship grant number PF6-170169 awarded by the Chandra X-ray Center, which is operated by the Smithsonian Astrophysical Observatory for NASA under contract NAS8-03060. Support for program \#14574 was provided by NASA through a grant from the Space Telescope Science Institute, which is operated by the Association of Universities for Research in Astronomy, Inc., under NASA contract NAS 5-26555. This material is based upon work supported by the National Science Foundation under Grant No. 1909203. E.R.-R. and R.W.E. thank the David and Lucile Packard Foundation, the Heising-Simons Foundation, and the Danish National Research Foundation (DNRF132) for support. R.W.E. is supported by the Eugene V. Cota-Robles Fellowship and National Science Foundation Graduate Research Fellowship Program (Award \#1339067). A.A. is supported by the Berkeley Graduate Fellowship and the Cranor Fellowship. Resources supporting this work were provided by the NASA High-End Computing (HEC) Program through the NASA Advanced Supercomputing (NAS) Division at Ames Research Center, by the Institute for Advanced Study, by the University of Copenhagen high-performance computing cluster funded by a grant from VILLUM FONDEN (project number 16599), and by Syracuse University. Any opinions, findings, and conclusions or recommendations expressed in this material are those of the authors and do not necessarily reflect the views of the National Science Foundation.}

\software{
FLASH~\citep{Fryxell2000}, yt~\citep{yt:2011}, Astropy~\citep{astropy:2013,astropy:2018}, Plotly~\citep{plotly}, Matplotlib~\citep{Hunter:2007}
}

\clearpage
\appendix
\section{Fitting Formulae To Coefficients of Drag and Accretion}\label{subsec:fits}
We present fitting formulae for the coefficients of accretion $C_{\mathrm a}$ and drag force $C_{\mathrm d}$ as a function of the mass ratio $q$ and upstream Mach number $\mathcal{M}_\infty$ from both our $\gamma = 4/3$ and $\gamma = 5/3$ simulations. Fits are constructed using the $q$, $\mathcal{M}_\infty$, $C_{\mathrm a}$, $C_{\mathrm d}$ datasets presented in Tables. \ref{tab:sims_43_params} and \ref{tab:sims_53_params} in Sec.~\ref{sec:hydro_sims}. The fits show a mapping between the simulation results to the input parameters for the parameter space we have explored. For the $\gamma = 4/3$ simulations, we use third order polynomials as fitting functions for both log$_{10}C_{\mathrm a}$ and log$_{10}C_{\mathrm d}$, that are  expressed as follows
\begin{equation}
\label{eq:logmdot_g43}
\begin{split}
\log_{10} C_{\mathrm a}^{(4/3)} = a_1^{(4/3)} + a_2^{(4/3)} q_{\rm r} + a_3^{(4/3)}\mathcal{M}_\infty + a_4^{(4/3)} q\mathcal{M}_\infty
+ a_5^{(4/3)} q_{\rm r}^2 + a_6^{(4/3)} \mathcal{M}_\infty^2 
+ a_7^{(4/3)} q_{\rm r} \mathcal{M}_\infty^2 \\ + a_8^{(4/3)} q_{\rm r}^2\mathcal{M}_\infty + a_9^{(4/3)} q_{\rm r}^3 + a_{10}^{(4/3)} \mathcal{M}_\infty^3
\end{split}
\end{equation}

\begin{equation}
\label{eq:logdrag_g43}
\begin{split}
\log_{10} C_{\mathrm d}^{(4/3)} = d_1^{(4/3)} + d_2^{(4/3)} q_{\rm r} + d_3^{(4/3)}\mathcal{M}_\infty + d_4^{(4/3)} q_{\rm r}\mathcal{M}_\infty + d_5^{(4/3)} q_{\rm r}^2 + d_6^{(4/3)} \mathcal{M}_\infty^2
+ d_7^{(4/3)} q_{\rm r}\mathcal{M}_\infty^2 \\ + d_8^{(4/3)} q_{\rm r}^2\mathcal{M}_\infty + d_9^{(4/3)} q_{\rm r}^3 + d_{10}^{(4/3)} \mathcal{M}_\infty^3
\end{split}
\end{equation}
                 
The least square solutions we obtain for the log$_{10} C_{\mathrm a}^{4/3}$ polynomial fit are $a_1^{(4/3)} = 0.816875242$, $a_2^{(4/3)} = -9.97843860$, $a_3^{(4/3)} = 0.138186031$, $a_4^{(4/3)} = -0.480338859$, $a_5^{(4/3)} = 46.9754907$, $a_6^{(4/3)} = -0.333034058$, $a_7^{(4/3)} = 0.671283731$, $a_8^{(4/3)} = -4.16198903$, $a_9^{(4/3)} = -58.9379466$, $a_{10}^{(4/3)} = 0.0378957450$. 
For fitting log$_{10} C_{\mathrm d}^{(4/3)}$, we have obtained $d_1^{(4/3)} = 0.551005528$, $d_2^{(4/3)} = 0.450225376$, $d_3^{(4/3)} = -0.674076156$, $d_4^{(4/3)} = 0.594634757$, $d_5^{(4/3)} = -14.9500309$, $d_6^{(4/3)} = 0.315907987$, $d_7^{(4/3)} = -0.0203304265$, $d_8^{(4/3)} = -1.70433247$, $d_9^{(4/3)} = 30.4494171$, $d_{10}^{(4/3)} = -0.0309267276$.

\begin{figure*}
  \centering
  \includegraphics[width=13cm]{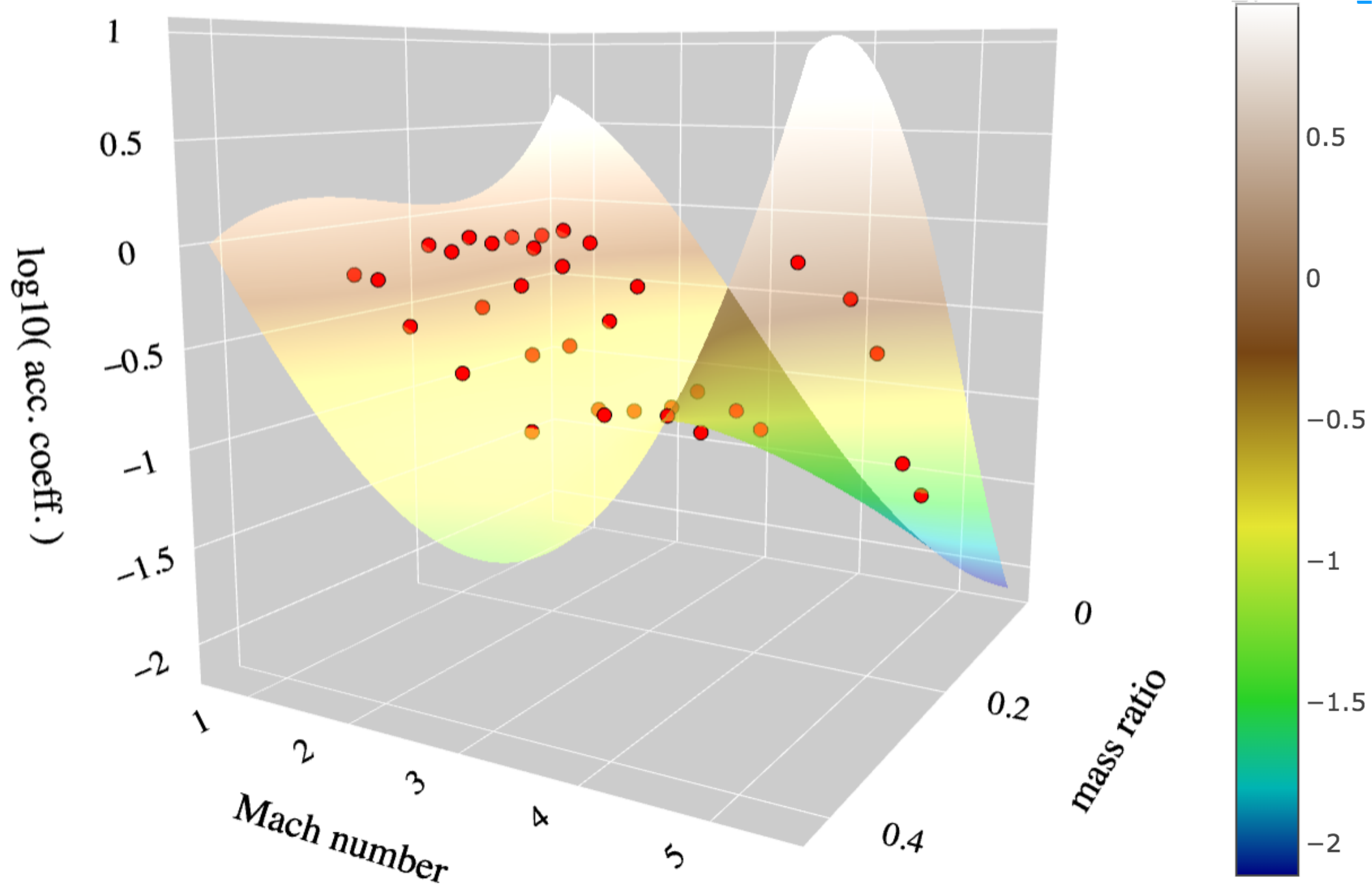}
\caption{Relation between the coefficient of accretion, mass ratio, and upstream Mach number---$\log_{10}C_{\mathrm a} (q_{\rm r}, \mathcal{M}_\infty)$ for $(\Gamma, \gamma) = (4/3, 4/3)$ flows. The red dots represent the $\log_{10}C_{\mathrm a}$ results obtained from the hydrodynamic simulations with $q_{\rm r}$ and $\mathcal{M}_\infty$ parameters. The three-dimensional surface shows the best fitting third-order polynomial relation of $\log_{10}C_{\mathrm a}$ in terms of $(q_{\rm r}, \mathcal{M}_\infty)$.\label{fig:logmdot_g43}}
\end{figure*}

\begin{figure*}
  \centering
  \includegraphics[width=13cm]{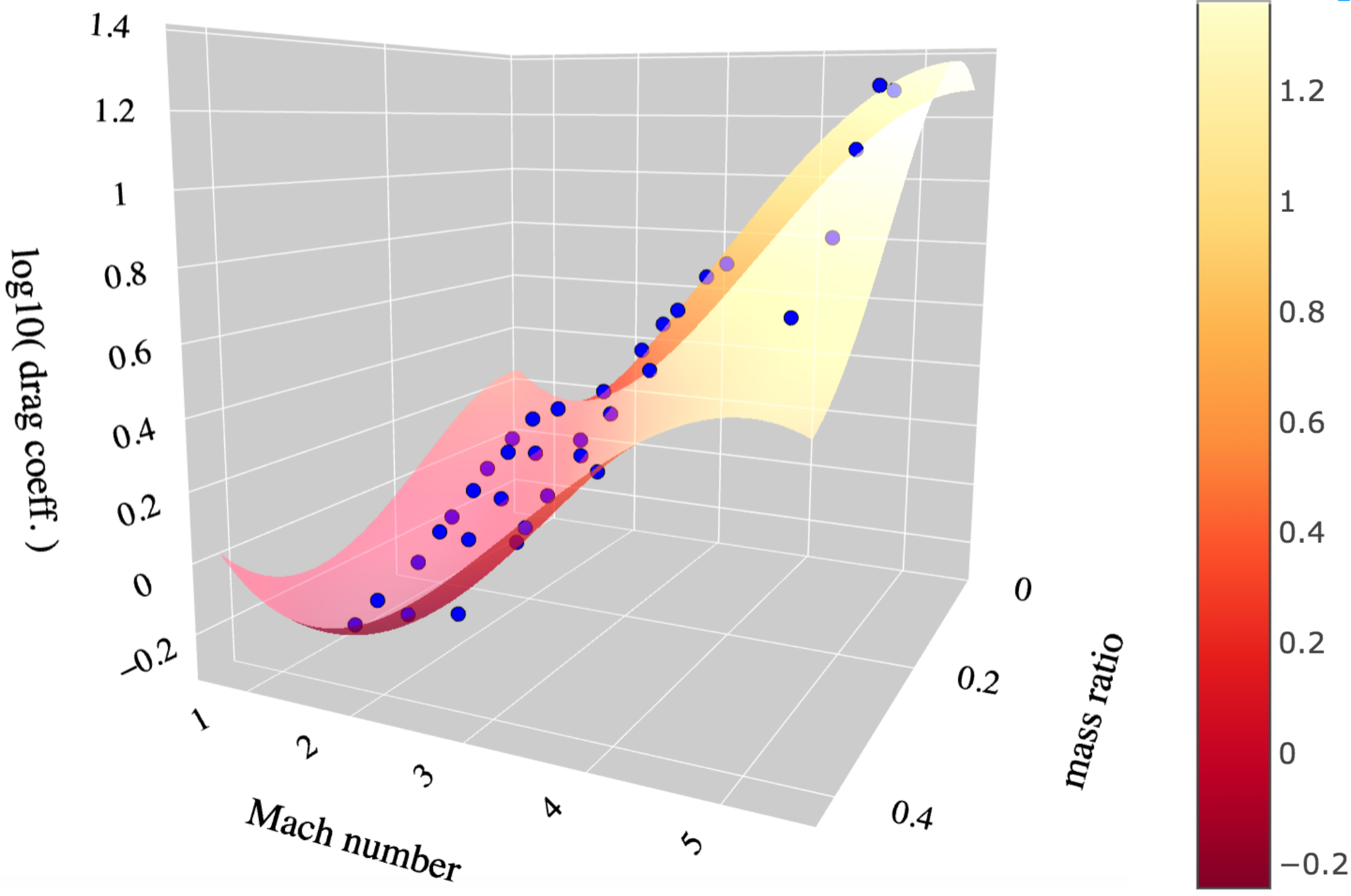}
\caption{Relation between the coefficient of drag, mass ratio, and upstream Mach number---$\log_{10}C_{\mathrm d} (q_{\rm r}, \mathcal{M}_\infty)$ for $(\Gamma, \gamma) = (4/3, 4/3)$ flows. The blue dots represent the $\log_{10}C_{\mathrm d}$ results obtained from the hydrodynamic simulations with $q_{\rm r}$ and $\mathcal{M}_\infty$ parameters. The three-dimensional surface shows the best fitting third-order polynomial relation of $\log_{10}C_{\mathrm d}$ in terms of $(q_{\rm r}, \mathcal{M}_\infty)$.\label{fig:logdrag_g43}}
\end{figure*}

In Figs.~\ref{fig:logmdot_g43} and \ref{fig:logdrag_g43}, we present the $\log_{10}C_{\mathrm a} (q_{\rm r}, \mathcal{M}_\infty)$ and $\log_{10}C_{\mathrm d} (q_{\rm r}, \mathcal{M}_\infty)$ datasets respectively from the $(\Gamma, \gamma) = (4/3, 4/3)$ simulations. Overlaid are the best fit $\log_{10}C_{\mathrm a} (q_{\rm r}, \mathcal{M}_\infty)$ and $\log_{10}C_{\mathrm d} (q_{\rm r}, \mathcal{M}_\infty)$ surfaces as presented in Eqns.~\ref{eq:logmdot_g43} and \ref{eq:logdrag_g43} above. 

Accumulation of material from accretion flows onto an embedded compact object requires either that the object be a black hole, or the presence of an effective cooling channel if the object has a surface. In the case of objects with a surface, accretion releases gravitational potential energy and generates feedback. Our simulations include a completely absorbing central boundary condition, and therefore our setup is appropriate for calculating accretion rates for cases where a mass accumulation from accretion is possible. The fitting formulae from the $\gamma = 4/3$ simulations presented above are applicable for systems where a black hole is inspiraling inside the envelope of a more massive giant branch star. This is because, taking into account the minimum mass of black holes and the fact that the envelope would be of a more massive giant star than the embedded object, the mass of the giant star in this scenario would be greater than $\sim 10 M_\odot$. As mentioned earlier, the flow of material in such high mass stars would be represented by a $\gamma = 4/3$ equation of state.  

For the $\gamma=5/3$ simulations, we use second order polynomials as fitting functions for both log$_{10}C_{\mathrm a}$  and log$_{10}C_{\mathrm d}$, which is expressed as

\begin{equation}
\label{eq:logmdot_g53}
\begin{split}
\log_{10} C_{\mathrm a}^{(5/3)} = a_1^{(5/3)} + a_2^{(5/3)} q_{\rm r} + a_3^{(5/3)}\mathcal{M}_\infty + a_4^{(5/3)} q_{\rm r}\mathcal{M}_\infty + a_5^{(5/3)} q_{\rm r}^2 + a_6^{(5/3)} \mathcal{M}_\infty^2
\end{split}
\end{equation}

\begin{equation}
\label{eq:logdrag_g53}
\begin{split}
\log_{10} C_{\mathrm d}^{(5/3)} = d_1^{(5/3)} + d_2^{(5/3)} q_{\rm r} + d_3^{(5/3)}\mathcal{M}_\infty + d_4^{(5/3)} q\mathcal{M}_\infty + d_5^{(5/3)} q_{\rm r}^2 + d_6^{(5/3)} \mathcal{M}_\infty^2
\end{split}
\end{equation}

The least square solutions we obtain for the log$_{10} C_{\mathrm a}^{5/3}$ polynomial fit are $a_1^{(5/3)} = 0.91841162$, $a_2^{(5/3)} = -0.96187412$, $a_3^{(5/3)} = -1.20569095$, $a_4^{(5/3)} = 1.22475534$, $a_5^{(5/3)} = -2.48004138$, $a_6^{(5/3)} = 0.1150233$. For fitting $\log_{10} C_{\mathrm d}^{(5/3)}$, we obtain $d_1^{(5/3)} = -0.15515258$, $d_2^{(5/3)} = -3.03230504$, $d_3^{(5/3)} = 0.27564942$, $d_4^{(5/3)} = 0.19765314$, $d_5^{(5/3)} = 1.41864156$, $d_6^{(5/3)} = 
-0.0092128$. 

\begin{figure*}
  \centering
  \includegraphics[width=13cm]{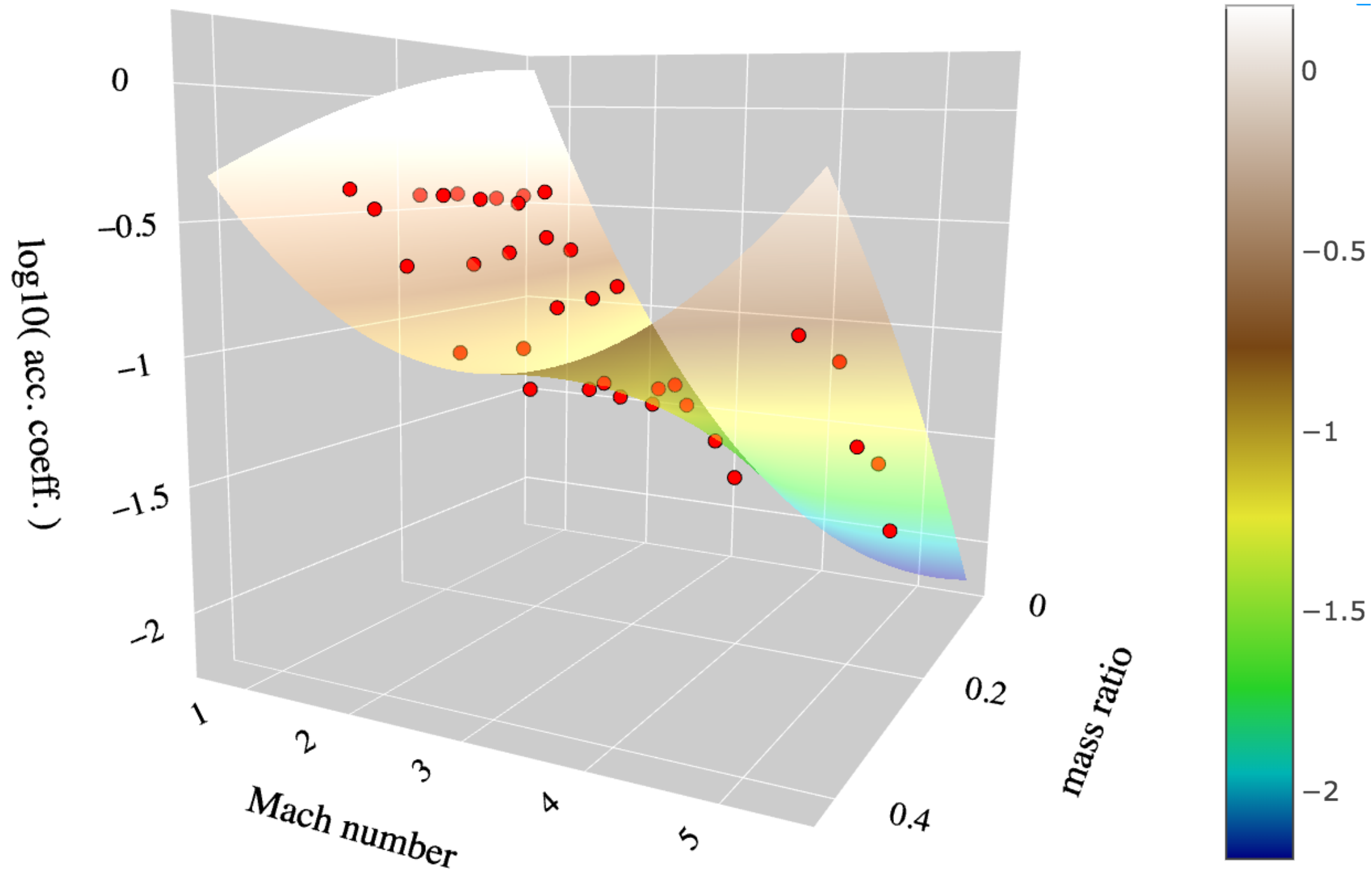}
\caption{Relation between the coefficient of accretion, mass ratio, and upstream Mach number---$\log_{10}C_{\mathrm a} (q_{\rm r}, \mathcal{M})$ for $(\Gamma, \gamma) = (5/3, 5/3)$ flows. The red dots represent the $\log_{10}C_{\mathrm a}$ results obtained from the hydrodynamic simulations with $q_{\rm r}$ and $\mathcal{M}_\infty$ parameters. The three-dimensional surface shows the best fitting second-order polynomial relation of $\log_{10}C_{\mathrm a}$ in terms of $(q_{\rm r}, \mathcal{M}_\infty)$.\label{fig:logmdot_g53}}.
\end{figure*}

\begin{figure*}
 \centering
 \includegraphics[width=13cm]{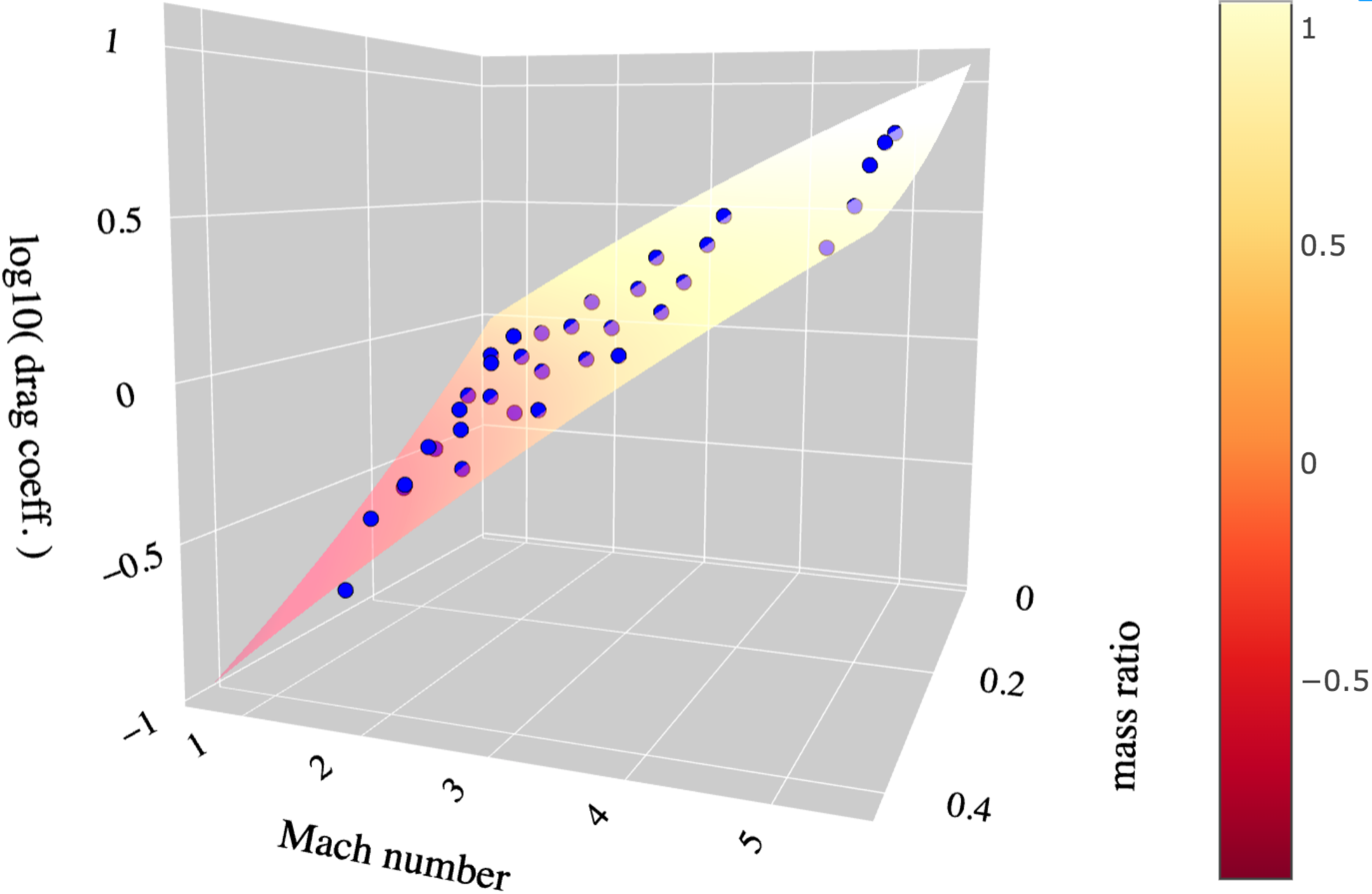}\caption{Relation between the coefficient of drag, mass ratio, and upstream Mach number---$\log_{10}C_{\mathrm d} (q_{\rm r}, \mathcal{M}_\infty)$ for $(\Gamma, \gamma) = (5/3, 5/3)$ flows. The blue dots represent the $\log_{10}C_{\mathrm d}$ results obtained from the hydrodynamic simulations with $q_{\rm r}$ and $\mathcal{M}_\infty$ parameters. The three-dimensional surface shows the best fitting second-order polynomial relation of $\log_{10}C_{\mathrm d}$ in terms of $(q_{\rm r}, \mathcal{M}_\infty)$.\label{fig:logdrag_g53}}.
\end{figure*}

In Figs.~\ref{fig:logmdot_g53} and \ref{fig:logdrag_g53}, we present the $\log_{10}C_{\mathrm a} (q_{\rm r}, \mathcal{M}_\infty)$ and $\log_{10}C_{\mathrm d} (q_{\rm r}, \mathcal{M}_\infty)$ datasets respectively from the $(\Gamma, \gamma) = (5/3, 5/3)$ simulations. Overlaid are the best fit $\log_{10}C_{\mathrm a} (q_{\rm r}, \mathcal{M}_\infty)$ and $\log_{10}C_{\mathrm d} (q_{\rm r}, \mathcal{M}_\infty)$ surfaces as presented in Eqns.~\ref{eq:logmdot_g53} and \ref{eq:logdrag_g53} above. These fitting formulae from the $\gamma = 5/3$ simulations are applicable for systems where a white dwarf or a main sequence star is inspiraling inside the envelope of a more massive giant branch star. The giant star in this case would be less massive than that in the $\gamma = 4/3$ systems. However, despite flow convergence in such systems, we do not expect significant mass accumulation from accretion onto the compact object due to the lack of an apparent cooling mechanism. Main sequence stars and white dwarfs are not compact enough to promote cooling channels such as neutrino emission, mediating the luminosity of the accretion onto the neutron stars. Also, the common envelope flows are optically thick, preventing the escape of heat through photon diffusion. It would be more appropriate to model these objects with a hard-surface boundary condition than an absorbing boundary condition. Interactive versions of Figures \ref{fig:logmdot_g43}, \ref{fig:logdrag_g43}, \ref{fig:logmdot_g53}, and \ref{fig:logdrag_g53} can be viewed at \verb"https://soumide1102.github.io/common-envelope-hydro-paper".

\begin{thebibliography}{}
\expandafter\ifx\csname natexlab\endcsname\relax\def\natexlab#1{#1}\fi
\providecommand{\url}[1]{\href{#1}{#1}}

\bibitem[{Aasi {et~al.}(2015)}]{TheLIGOScientific:2014jea}
Aasi, J., {et~al.} 2015, Class. Quant. Grav., 32, 074001

\bibitem[{Abbott {et~al.}(2016{\natexlab{a}})}]{Abbott:2016blz}
Abbott, B.~P., {et~al.} 2016{\natexlab{a}}, Phys. Rev. Lett., 116, 061102

\bibitem[{Abbott {et~al.}(2016{\natexlab{b}})}]{TheLIGOScientific:2016pea}
---. 2016{\natexlab{b}}, Phys. Rev., X6, 041015

\bibitem[{Abbott {et~al.}(2017{\natexlab{a}})}]{Abbott:2017vtc}
---. 2017{\natexlab{a}}, Phys. Rev. Lett., 118, 221101

\bibitem[{Abbott {et~al.}(2017{\natexlab{b}})}]{Abbott:2017gyy}
---. 2017{\natexlab{b}}, Astrophys. J., 851, L35

\bibitem[{Abbott {et~al.}(2017{\natexlab{c}})}]{Abbott:2017oio}
---. 2017{\natexlab{c}}, Phys. Rev. Lett., 119, 141101

\bibitem[{Abbott {et~al.}(2019)}]{LIGOScientific:2018mvr}
---. 2019, Phys. Rev., X9, 031040

\bibitem[{Acernese {et~al.}(2015)}]{TheVirgo:2014hva}
Acernese, F., {et~al.} 2015, Class. Quant. Grav., 32, 024001

\bibitem[{{Antoni} {et~al.}(2019){Antoni}, {MacLeod}, \&
  {Ramirez-Ruiz}}]{Antoni:2019pgq}
{Antoni}, A., {MacLeod}, M., \& {Ramirez-Ruiz}, E. 2019, \apj, 884, 22

\bibitem[{{Astropy Collaboration}(2013)}]{astropy:2013}
{Astropy Collaboration}. 2013, \aap, 558, A33

\bibitem[{{Astropy Collaboration}(2018)}]{astropy:2018}
---. 2018, \aj, 156, 123

\bibitem[{{Bailyn} {et~al.}(1998){Bailyn}, {Jain}, {Coppi}, \&
  {Orosz}}]{Bailyn:1998}
{Bailyn}, C.~D., {Jain}, R.~K., {Coppi}, P., \& {Orosz}, J.~A. 1998, \apj, 499,
  367

\bibitem[{Batta \& Ramirez-Ruiz(2019)}]{Batta:2019clm}
Batta, A., \& Ramirez-Ruiz, E. 2019, arXiv:1904.04835

\bibitem[{Bavera {et~al.}(2019)Bavera, Fragos, Qin, Zapartas, Neijssel, Mandel,
  Batta, Gaebel, Kimball, \& Stevenson}]{Bavera:2019fkg}
Bavera, S.~S., Fragos, T., Qin, Y., {et~al.} 2019, arXiv:1906.12257

\bibitem[{{Belczynski} {et~al.}(2014){Belczynski}, {Buonanno}, {Cantiello},
  {Fryer}, {Holz}, {Mandel}, {Miller}, \& {Walczak}}]{Belczynski:2014}
{Belczynski}, K., {Buonanno}, A., {Cantiello}, M., {et~al.} 2014, \apj, 789,
  120

\bibitem[{Belczynski {et~al.}(2016)Belczynski, Holz, Bulik, \&
  O'Shaughnessy}]{Belczynski:2016obo}
Belczynski, K., Holz, D.~E., Bulik, T., \& O'Shaughnessy, R. 2016, Nature, 534,
  512

\bibitem[{{Bethe} \& {Brown}(1998)}]{Bethe:1998}
{Bethe}, H.~A., \& {Brown}, G.~E. 1998, \apj, 506, 780

\bibitem[{Biwer {et~al.}(2019)Biwer, Capano, De, Cabero, Brown, Nitz, \&
  Raymond}]{Biwer:2018osg}
Biwer, C.~M., Capano, C.~D., De, S., {et~al.} 2019, Publ. Astron. Soc. Pac.,
  131, 024503

\bibitem[{{Blondin} \& {Raymer}(2012)}]{2012ApJ...752...30B}
{Blondin}, J.~M., \& {Raymer}, E. 2012, \apj, 752, 30

\bibitem[{{Bondi} \& {Hoyle}(1944)}]{1944MNRAS.104..273B}
{Bondi}, H., \& {Hoyle}, F. 1944, \mnras, 104, 273

\bibitem[{{Brown}(1995)}]{Brown:1995}
{Brown}, G.~E. 1995, \apj, 440, 270

\bibitem[{Chamandy {et~al.}(2019)Chamandy, Blackman, Frank, Carroll-Nellenback,
  Zou, \& Tu}]{Chamandy:2019psk}
Chamandy, L., Blackman, E.~G., Frank, A., {et~al.} 2019, arXiv:1908.06195

\bibitem[{{Chamandy} {et~al.}(2019){Chamandy}, {Tu}, {Blackman},
  {Carroll-Nellenback}, {Frank}, {Liu}, \& {Nordhaus}}]{Chamandy:2018a}
{Chamandy}, L., {Tu}, Y., {Blackman}, E.~G., {et~al.} 2019, \mnras, 486, 1070

\bibitem[{{Chamandy} {et~al.}(2018){Chamandy}, {Frank}, {Blackman},
  {Carroll-Nellenback}, {Liu}, {Tu}, {Nordhaus}, {Chen}, \&
  {Peng}}]{Chamandy:2018}
{Chamandy}, L., {Frank}, A., {Blackman}, E.~G., {et~al.} 2018, \mnras, 480,
  1898

\bibitem[{{Chandrasekhar}(1943)}]{1943ApJ....97..255C}
{Chandrasekhar}, S. 1943, \apj, 97, 255

\bibitem[{{Chevalier}(1993)}]{Chevalier:1993}
{Chevalier}, R.~A. 1993, \apjl, 411, L33

\bibitem[{De {et~al.}(2018)De, Biwer, Capano, Nitz, \& Brown}]{De:2018zrk}
De, S., Biwer, C.~M., Capano, C.~D., Nitz, A.~H., \& Brown, D.~A. 2018,
  arXiv:1811.09232

\bibitem[{{De Marco} \& {Izzard}(2017)}]{2017PASA...34....1D}
{De Marco}, O., \& {Izzard}, R.~G. 2017, \pasa, 34, e001

\bibitem[{Edgar(2004)}]{Edgar:2004}
Edgar, R. 2004, New Astronomy Reviews, 48, 843

\bibitem[{{Eldridge} \& {Stanway}(2016)}]{EldridgeStanway:2016}
{Eldridge}, J.~J., \& {Stanway}, E.~R. 2016, \mnras, 462, 3302

\bibitem[{Everson {et~al.}(2019)Everson, MacLeod, De, Macias, \&
  Ramirez-Ruiz}]{Rosa:2019}
Everson, R.~W., MacLeod, M., De, S., Macias, P., \& Ramirez-Ruiz, E. 2019

\bibitem[{Farr {et~al.}(2018)Farr, Holz, \& Farr}]{Farr:2017gtv}
Farr, B., Holz, D.~E., \& Farr, W.~M. 2018, Astrophys. J., 854, L9

\bibitem[{{Farr} {et~al.}(2011){Farr}, {Sravan}, {Cantrell}, {Kreidberg},
  {Bailyn}, {Mandel}, \& {Kalogera}}]{Farr:2011}
{Farr}, W.~M., {Sravan}, N., {Cantrell}, A., {et~al.} 2011, \apj, 741, 103

\bibitem[{Farr {et~al.}(2017)Farr, Stevenson, Coleman~Miller, Mandel, Farr, \&
  Vecchio}]{Farr:2017uvj}
Farr, W.~M., Stevenson, S., Coleman~Miller, M., {et~al.} 2017, Nature, 548, 426

\bibitem[{Fragos {et~al.}(2019)Fragos, Andrews, Ramirez-Ruiz, Meynet, Kalogera,
  Taam, \& Zezas}]{Fragos:2019box}
Fragos, T., Andrews, J.~J., Ramirez-Ruiz, E., {et~al.} 2019, Astrophys. J.,
  883, L45

\bibitem[{Fryxell \& Taam(1989)}]{Fryxell:1988}
Fryxell, B., \& Taam, R. 1989, The Astrophysical Journal, 335,
  doi:10.1086/166973

\bibitem[{Fryxell {et~al.}(1987)Fryxell, Taam, \& McMillan}]{Fryxell:1987}
Fryxell, B., Taam, R., \& McMillan, S. 1987, The Astrophysical Journal, 315,
  doi:10.1086/165157

\bibitem[{Fryxell {et~al.}(2000)Fryxell, Olson, Ricker, Timmes, Zingale, Lamb,
  MacNeice, Rosner, Truran, \& Tufo}]{Fryxell2000}
Fryxell, B., Olson, K., Ricker, P., {et~al.} 2000, Astrophysical Journal,
  Supplement, 131, 273

\bibitem[{Fuller \& Ma(2019)}]{Fuller:2019sxi}
Fuller, J., \& Ma, L. 2019, Astrophys. J., 881, L1

\bibitem[{Han {et~al.}(1994)Han, Podsiadlowski, \& Eggleton}]{Han:1994}
Han, Z., Podsiadlowski, P., \& Eggleton, P.~P. 1994, Monthly Notices of the
  Royal Astronomical Society, 270, 121

\bibitem[{Han {et~al.}(2002)Han, Podsiadlowski, Maxted, Marsh, \&
  Ivanova}]{Han:2002}
Han, Z., Podsiadlowski, P., Maxted, P. F.~L., Marsh, T.~R., \& Ivanova, N.
  2002, Monthly Notices of the Royal Astronomical Society, 336, 449

\bibitem[{{Hoyle} \& {Lyttleton}(1939)}]{1939PCPS...35..405H}
{Hoyle}, F., \& {Lyttleton}, R.~A. 1939, Proceedings of the Cambridge
  Philosophical Society, 35, 405

\bibitem[{Hunter(2007)}]{Hunter:2007}
Hunter, J.~D. 2007, Computing in Science \& Engineering, 9, 90

\bibitem[{{Iaconi} {et~al.}(2018{\natexlab{a}}){Iaconi}, {De Marco}, {Passy},
  \& {Staff}}]{Iaconi:2018}
{Iaconi}, R., {De Marco}, O., {Passy}, J.-C., \& {Staff}, J.
  2018{\natexlab{a}}, \mnras, 477, 2349

\bibitem[{{Iaconi} {et~al.}(2018{\natexlab{b}}){Iaconi}, {De Marco}, {Passy},
  \& {Staff}}]{2018MNRAS.477.2349I}
---. 2018{\natexlab{b}}, \mnras, 477, 2349

\bibitem[{{Iaconi} {et~al.}(2017){Iaconi}, {Reichardt}, {Staff}, {De Marco},
  {Passy}, {Price}, {Wurster}, \& {Herwig}}]{Iaconi:2017}
{Iaconi}, R., {Reichardt}, T., {Staff}, J., {et~al.} 2017, \mnras, 464, 4028

\bibitem[{{Iben} \& {Livio}(1993)}]{1993PASP..105.1373I}
{Iben}, Icko, J., \& {Livio}, M. 1993, \pasp, 105, 1373

\bibitem[{{Ivanova} \& {Nandez}(2016)}]{Ivanova:2016}
{Ivanova}, N., \& {Nandez}, J.~L.~A. 2016, \mnras, 462, 362

\bibitem[{{Ivanova} {et~al.}(2013){Ivanova}, {Justham}, {Chen}, {De Marco},
  {Fryer}, {Gaburov}, {Ge}, {Glebbeek}, {Han}, {Li}, {Lu}, {Marsh},
  {Podsiadlowski}, {Potter}, {Soker}, {Taam}, {Tauris}, {van den Heuvel}, \&
  {Webbink}}]{2013A&ARv..21...59I}
{Ivanova}, N., {Justham}, S., {Chen}, X., {et~al.} 2013, \aapr, 21, 59

\bibitem[{King \& Kolb(1999)}]{kingkolb:1999}
King, A.~R., \& Kolb, U. 1999, Monthly Notices of the Royal Astronomical
  Society, 305, 654

\bibitem[{{Kreidberg} {et~al.}(2012){Kreidberg}, {Bailyn}, {Farr}, \&
  {Kalogera}}]{Kreidberg:2012}
{Kreidberg}, L., {Bailyn}, C.~D., {Farr}, W.~M., \& {Kalogera}, V. 2012, \apj,
  757, 36

\bibitem[{Kruckow {et~al.}(2016)Kruckow, Tauris, Langer, Szécsi, Marchant, \&
  Podsiadlowski}]{Kruckow:2016tti}
Kruckow, M.~U., Tauris, T.~M., Langer, N., {et~al.} 2016, Astron. Astrophys.,
  596, A58

\bibitem[{Kushnir {et~al.}(2016)Kushnir, Zaldarriaga, Kollmeier, \&
  Waldman}]{Kushnir:2016zee}
Kushnir, D., Zaldarriaga, M., Kollmeier, J.~A., \& Waldman, R. 2016, Mon. Not.
  Roy. Astron. Soc., 462, 844

\bibitem[{{Lucy}(1967)}]{Lucy:1967}
{Lucy}, L.~B. 1967, \aj, 72, 813

\bibitem[{{MacLeod} {et~al.}(2017){MacLeod}, {Antoni}, {Murguia-Berthier},
  {Macias}, \& {Ramirez-Ruiz}}]{MacLeod:2017}
{MacLeod}, M., {Antoni}, A., {Murguia-Berthier}, A., {Macias}, P., \&
  {Ramirez-Ruiz}, E. 2017, \apj, 838, 56

\bibitem[{MacLeod \& Ramirez-Ruiz(2015{\natexlab{a}})}]{MacLeod_2015}
MacLeod, M., \& Ramirez-Ruiz, E. 2015{\natexlab{a}}, The Astrophysical Journal,
  803, 41

\bibitem[{MacLeod \& Ramirez-Ruiz(2015{\natexlab{b}})}]{MacLeod:2014yda}
---. 2015{\natexlab{b}}, Astrophys. J., 798, L19

\bibitem[{Mandel \& Farmer(2018)}]{Mandel:2018hfr}
Mandel, I., \& Farmer, A. 2018, arXiv:1806.05820

\bibitem[{{Mapelli}(2018)}]{Mapelli:2018}
{Mapelli}, M. 2018, arXiv e-prints, arXiv:1809.09130

\bibitem[{{Marchant} {et~al.}(2016){Marchant}, {Langer}, {Podsiadlowski},
  {Tauris}, \& {Moriya}}]{Marchant:2016}
{Marchant}, P., {Langer}, N., {Podsiadlowski}, P., {Tauris}, T.~M., \&
  {Moriya}, T.~J. 2016, \aap, 588, A50

\bibitem[{Murguia-Berthier {et~al.}(2017)Murguia-Berthier, MacLeod,
  Ramirez-Ruiz, Antoni, \& Macias}]{Murguia-Berthier:2017}
Murguia-Berthier, A., MacLeod, M., Ramirez-Ruiz, E., Antoni, A., \& Macias, P.
  2017, The Astrophysical Journal, 845, 173

\bibitem[{Nandez {et~al.}(2015)Nandez, Ivanova, \& Lombardi}]{Nandez:2015}
Nandez, J. L.~A., Ivanova, N., \& Lombardi, J.~C., J. 2015, Monthly Notices of
  the Royal Astronomical Society: Letters, 450, L39

\bibitem[{Nitz {et~al.}(2019)Nitz, Capano, Nielsen, Reyes, White, Brown, \&
  Krishnan}]{Nitz:2018imz}
Nitz, A.~H., Capano, C., Nielsen, A.~B., {et~al.} 2019, Astrophys. J., 872, 195

\bibitem[{{Ohlmann} {et~al.}(2016{\natexlab{a}}){Ohlmann}, {R{\"o}pke},
  {Pakmor}, \& {Springel}}]{Ohlmann:2016b}
{Ohlmann}, S.~T., {R{\"o}pke}, F.~K., {Pakmor}, R., \& {Springel}, V.
  2016{\natexlab{a}}, \apjl, 816, L9

\bibitem[{{Ohlmann} {et~al.}(2016{\natexlab{b}}){Ohlmann}, {R{\"o}pke},
  {Pakmor}, {Springel}, \& {M{\"u}ller}}]{Ohlmann:2016a}
{Ohlmann}, S.~T., {R{\"o}pke}, F.~K., {Pakmor}, R., {Springel}, V., \&
  {M{\"u}ller}, E. 2016{\natexlab{b}}, \mnras, 462, L121

\bibitem[{{Ostriker}(1999)}]{1999ApJ...513..252O}
{Ostriker}, E.~C. 1999, \apj, 513, 252

\bibitem[{{{\"O}zel} {et~al.}(2010){{\"O}zel}, {Psaltis}, {Narayan}, \&
  {McClintock}}]{Ozel:2010}
{{\"O}zel}, F., {Psaltis}, D., {Narayan}, R., \& {McClintock}, J.~E. 2010,
  \apj, 725, 1918

\bibitem[{{Paczynski}(1976)}]{Paczynski:1976}
{Paczynski}, B. 1976, in IAU Symposium, Vol.~73, Structure and Evolution of
  Close Binary Systems, ed. P.~{Eggleton}, S.~{Mitton}, \& J.~{Whelan}, 75

\bibitem[{Passy {et~al.}(2011)Passy, Marco, Fryer, Herwig, Diehl, Oishi, Low,
  Bryan, \& Rockefeller}]{Passy_2011}
Passy, J.-C., Marco, O.~D., Fryer, C.~L., {et~al.} 2011, The Astrophysical
  Journal, 744, 52

\bibitem[{Piran \& Piran(2019)}]{Piran_2019}
Piran, Z., \& Piran, T. 2019, arXiv:1910.11358

\bibitem[{Plotly(2015)}]{plotly}
Plotly. 2015, Collaborative data science,  Montreal, QC: Plotly Technologies
  Inc.
\newblock \url{https://plot.ly}

\bibitem[{Ricker \& Taam(2007)}]{Ricker_2007}
Ricker, P.~M., \& Taam, R.~E. 2007, The Astrophysical Journal, 672, L41

\bibitem[{Ricker \& Taam(2012)}]{Ricker_2012}
---. 2012, The Astrophysical Journal, 746, 74

\bibitem[{Roxburgh(1967)}]{Roxburgh:1967}
Roxburgh, I.~W. 1967, Nature, 215, doi:10.1038/215838a0

\bibitem[{{Ruffert}(1994)}]{1994A&AS..106..505R}
{Ruffert}, M. 1994, \aaps, 106, 505

\bibitem[{{Ruffert}(1995)}]{1995A&AS..113..133R}
---. 1995, \aaps, 113, 133

\bibitem[{{Ruffert} \& {Arnett}(1994)}]{1994ApJ...427..351R}
{Ruffert}, M., \& {Arnett}, D. 1994, \apj, 427, 351

\bibitem[{Sandquist {et~al.}(1998)Sandquist, Taam, Lin, \&
  Burkert}]{Sandquist_1998}
Sandquist, E., Taam, R.~E., Lin, D. N.~C., \& Burkert, A. 1998, The
  Astrophysical Journal, 506, L65

\bibitem[{{Schr{\o}der} {et~al.}(2018){Schr{\o}der}, Batta, \&
  Ramirez-Ruiz}]{Schroder:2018hxk}
{Schr{\o}der}, S.~L., Batta, A., \& Ramirez-Ruiz, E. 2018, Astrophys. J., 862,
  L3

\bibitem[{Smarr \& Blandford(1976)}]{Smarr:1976}
Smarr, L.~L., \& Blandford, R. 1976, \apj, 207, 574

\bibitem[{{Stevenson} {et~al.}(2017){Stevenson}, {Vigna-G{\'o}mez}, {Mandel},
  {Barrett}, {Neijssel}, {Perkins}, \& {de Mink}}]{Stevenson:2017}
{Stevenson}, S., {Vigna-G{\'o}mez}, A., {Mandel}, I., {et~al.} 2017, Nature
  Communications, 8, 14906

\bibitem[{Taam \& Fryxell(1989)}]{Taam:1989}
Taam, R., \& Fryxell, B. 1989, The Astrophysical Journal, 339,
  doi:10.1086/167297

\bibitem[{{Taam} {et~al.}(1978){Taam}, {Bodenheimer}, \&
  {Ostriker}}]{1978ApJ...222..269T}
{Taam}, R.~E., {Bodenheimer}, P., \& {Ostriker}, J.~P. 1978, \apj, 222, 269

\bibitem[{{Taam} \& {Ricker}(2010)}]{2010NewAR..54...65T}
{Taam}, R.~E., \& {Ricker}, P.~M. 2010, \nar, 54, 65

\bibitem[{Tiwari {et~al.}(2018)Tiwari, Fairhurst, \& Hannam}]{Tiwari:2018qch}
Tiwari, V., Fairhurst, S., \& Hannam, M. 2018, Astrophys. J., 868, 140

\bibitem[{Turk {et~al.}(2011)Turk, Smith, Oishi, Skory, Skillman, Abel, \&
  Norman}]{yt:2011}
Turk, M.~J., Smith, B.~D., Oishi, J.~S., {et~al.} 2011, The Astrophysical
  Journal Supplement Series, 192, 9

\bibitem[{{van den Heuvel}(1976)}]{Heuvel:1976}
{van den Heuvel}, E.~P.~J. 1976, in IAU Symposium, Vol.~73, Structure and
  Evolution of Close Binary Systems, ed. P.~{Eggleton}, S.~{Mitton}, \&
  J.~{Whelan}, 35

\bibitem[{Venumadhav {et~al.}(2019)Venumadhav, Zackay, Roulet, Dai, \&
  Zaldarriaga}]{Venumadhav:2019lyq}
Venumadhav, T., Zackay, B., Roulet, J., Dai, L., \& Zaldarriaga, M. 2019,
  arXiv:1904.07214

\bibitem[{{Woosley}(2017)}]{Woosley:2017}
{Woosley}, S.~E. 2017, \apj, 836, 244

\bibitem[{{Yusof} {et~al.}(2013){Yusof}, {Hirschi}, {Meynet}, {Crowther},
  {Ekstr{\"o}m}, {Frischknecht}, {Georgy}, {Abu Kassim}, \&
  {Schnurr}}]{Yusof:2013}
{Yusof}, N., {Hirschi}, R., {Meynet}, G., {et~al.} 2013, \mnras, 433, 1114

\bibitem[{Zackay {et~al.}(2019)Zackay, Venumadhav, Dai, Roulet, \&
  Zaldarriaga}]{Zackay:2019tzo}
Zackay, B., Venumadhav, T., Dai, L., Roulet, J., \& Zaldarriaga, M. 2019, Phys.
  Rev., D100, 023007

\bibitem[{Zaldarriaga {et~al.}(2018)Zaldarriaga, Kushnir, \&
  Kollmeier}]{Zaldarriaga:2017qkw}
Zaldarriaga, M., Kushnir, D., \& Kollmeier, J.~A. 2018, Mon. Not. Roy. Astron.
  Soc., 473, 4174

\end{thebibliography}


\end{document}